\documentclass[12pt,preprint]{aastex}
\usepackage{lineno}
\newcommand       \Angstrom     {\,{\rm \AA}}

\newcommand       \cm           {\,{\rm cm}}

\newcommand       \pc           {\,{\rm pc}}

\newcommand       \nH           {n_{\rm H}}
\newcommand       \simlt        {\lesssim}
\newcommand       \simgt        {\gtrsim}
\newcommand       \gtsim        {\gtrsim}

\newcommand       \mum          {\,{\rm \mu m}}

\newcommand       \Msun         {\,{{\rm M}_\odot}}

\newcommand       \simali       {\sim\,}
\newcommand       \magni        {\,{\rm mag}}

\def    \xo             {x_{\rm o}}
\def    \cpone		{c_1^{\prime}}
\def    \cptwo		{c_2^{\prime}}
\def    \cpthree	{c_3^{\prime}}
\def    \cpfour	        {c_4^{\prime}}
\def    \cone		{c_1}
\def    \ctwo		{c_2}
\def    \cthree  	{c_3}
\def    \cfour	        {c_4}

\newcommand{\bigsigma}{\mbox{\Large \ensuremath{\sigma}}}
\countdef\decade=200
\advance\decade by \year
\countdef\hours=201
\advance\hours by \time
\divide\hours by 60
\countdef\mins=202
\advance\mins by \hours
\multiply\mins by 60
\multiply\hours by 100
\countdef\miltime=203
\advance\miltime by \hours
\advance\miltime by \time
\advance\miltime by -\mins
\def\today{\number\decade.\number\month.\number\day.\number\miltime}

\shorttitle{Dust at High Galactic Latitudes}
\title{
\vspace*{-2.0em}
{\normalsize\rm Accepted for publication in
               {\it The Astrophysical Journal}}\\
\vspace*{1.0em}
Ultraviolet Interstellar Extinction 
toward High Galactic Latitudes
\\{\small DRAFT: \today ~~}
}
\author{Ajay Mishra\altaffilmark{1,2}
             and Aigen Li\altaffilmark{2}}
\altaffiltext{1} {Department of Science, Technology, and Mathematics,
                  Lincoln University,
                  Jefferson City, MO 65101, USA;
                  {\sf MishraA@lincolnu.edu}
                   }
\altaffiltext{2}{Department of Physics and Astronomy, 
                 University of Missouri-Columbia, 
                 Columbia, MO, 65201, USA;
                 {\sf LiA@missouri.edu}
                 }

\begin{document}

\begin{abstract}
Nearby high Galactic latitude clouds provide
a unique laboratory to study the physics
and structure of the Galactic interstellar medium (ISM) 
and the properties of the interstellar dust.
In this work we select 32 sightlines toward reddened
background stars at high Galactic latitudes
with $|b|$\,$>$\,20$\degr$
for which high-quality spectra
from the {\it International Ultraviolet Explorer}
are available. We utilize the ``pair-method'' derive
the ultraviolet interstellar extinction curves
for these sightlines. We examine the extinction
properties of these sightlines and find no systematic
variations with the Galactic latitudes, although they
do show appreciable sightline-to-sightline variations.

%
\end{abstract}

\section{Introduction}\label{sec:intro}
The exact amount and nature of interstellar matter
at high Galactic latitudes is a subject of ongoing scrutiny.
High Galactic latitude clouds (HLCs) serve as
an ideal ``laboratory'' for studying the physics and
chemistry of the Galactic interstellar medium (ISM),
particularly at a stage before star formation.
With a distance of $\simali$150\,pc from the Earth
and $\simali$100\,pc from the Galactic plane, 
HLCs, as a group, are the nearest clouds to the Sun
and situated at the nearer edges of the Local Bubble
(e.g., see Magnani et al.\ 1985). 

Ranging in size from $\simlt$\,0.1 to $\simali$10$\pc$
and in mass from $\simali$\,0.1 to $\simali$1,000$\Msun$,
HLCs consist primarily of translucent clouds,
a transitional phase between diffuse and dense clouds
(see van Dishoeck \& Black 1988).
In diffuse clouds (e.g., sightlines toward $\zeta$ Oph
and $\zeta$ Per), hydrogen is mostly in atomic form
(H) and the gas-phase carbon is mostly in ionized form
(C$^{+}$). They typically have a visual extinction of
$A_V$\,$\simali$0--1$\magni$ and a hydrogen
number density of $\nH$\,$\simali$100$\cm^{-3}$. 
Dense clouds are subject to large extinction
(with $A_V$\,$>$\,5$\magni$) and their
densities typically exceed $10^4\cm^{-3}$.
In dense clouds, most of the gas-phase carbon
becomes tied up in the form of CO. 
Translucent clouds represent the transitional regime 
between diffuse clouds, in which chemistry is driven
primarily by photoprocesses, and dense clouds,
in which collisional processes dominate the reaction network.
In translucent clouds, hydrogen is mostly molecular
(i.e., H$_2$) and the transition of carbon from ionized
to atomic (C) or molecular (CO) form takes place.
Indeed, unlike diffuse clouds, translucent clouds
are easily detectable in the CO J\,=\,1--0 transition
(e.g., see Jiang et al.\ 2026, Zhang et al.\ 2026).
With a typical visual extinction of
$A_V$\,$\simali$1--2$\magni$, translucent clouds 
are ``translucent'' to the interstellar radiation field,
i.e., ultraviolet (UV) starlight can penetrate into the clouds.

The unique properties of HLCs (i.e., translucent in nature)
and their proximity (so that they can be observed with
high spatial resolution) as well as their extraplanar
locations (so that they can be observed with less
confusion than clouds in the Galactic plane),
provide an important opportunity to gain insight
into the nature of the local ISM.
Moreover, a close correlation exists
between the large-scale, extended filamentary
``cirrus'' infrared (IR) emission at 60 and 100$\mum$
(Low et al.\ 1984) observed with
the {\it Infrared Astronomical Satellite} (IRAS)
and the CO emission from HLCs
(Blitz et al.\ 1984, Magnani et al.\ 1985).
These HLCs are identified with the cores of
the more extended IR cirrus discovered by IRAS
(Weiland et al. 1986). This association also firmly
establishes HLCs as an important agent for studying
the nature, origin and evolution as well as the regional
variation of interstellar dust at high Galactic latitudes.

The IRAS observations of HLCs at 12, 25, 60 and 100$\mum$
have revealed significant cloud-to-cloud variations
(Weiland et al.\ 1986, Verter et al.\ 2000).
Due to their translucent nature, dust grains in HLCs
are heated relatively uniformly by the local interstellar
radiation field. The interstellar radiation field penetrates
translucent clouds  thoroughly, ensuring nearly uniform
heating. This is also because HLCs show no sign
of star formation (see Magnani et al.\ 1995)
so that the external radiation field is
the dominant heating source for these clouds.
Therefore, the cloud-to-cloud variations in IR emission
observed by IRAS represent differences in grain contents
(i.e., dust sizes and abundances). This makes HLCs ideal
for exploring the dust properties of the local ISM.

In addition the thermal IR emission, the properties of dust
(e.g., sizes, compositions, and abundances) can also be
constrained by their absorption and scattering
(the combination of which is called extinction)
of starlight in the UV and visible
(as well as the near- and mid-IR) wavelengths.
To appreciate the importance of HLCs
as a unique class of interstellar clouds, 
we launch a project to systematically study
the properties of dust at high Galactic latitudes.
In this work, we derive the UV exinction curves
for 32 HLCs. In \S\ref{sec:sample} we describe
how the sample is selected and how the UV spectra
of the sample sources are constructed.
In \S\ref{sec:extcurv} we derive the UV extinction
curves for all the selected HLCs.
The results are discussed in \S\ref{sec:discussion}
and summarized in \S\ref{sec:summary}.



%

\section{The Sample}\label{sec:sample}
\begin{figure*}
\centering
\includegraphics[width=0.9\textwidth]{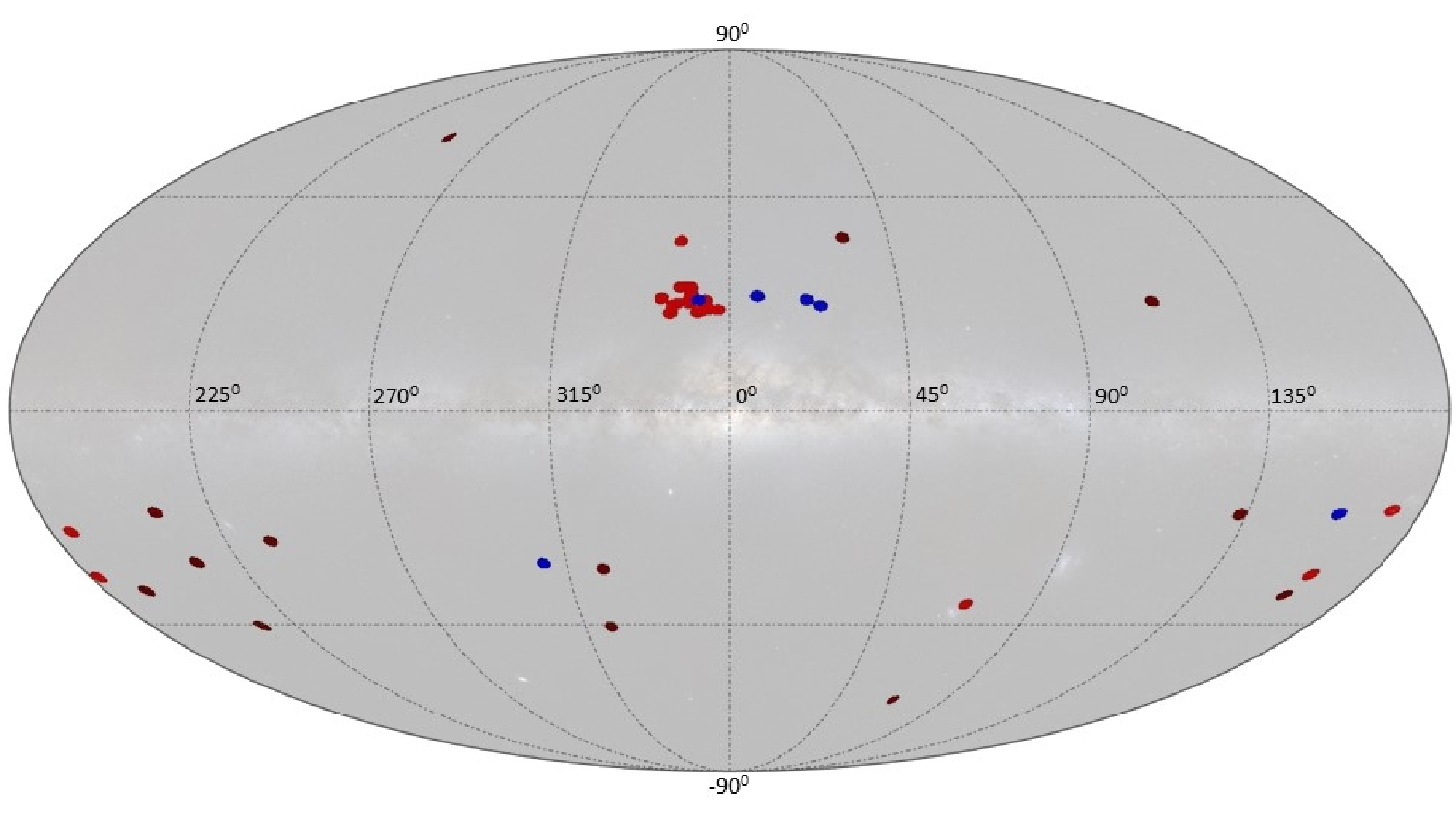}
\vspace{-2mm}
\caption{\footnotesize
  \label{fig:map}
  The distribution of the selected high latitude sources
  with $|b|\simgt20^{\degr}$ projected on the Galactic
  longitude–latitude plane.
  Red and violet circles represent reddenned stars,
  while blue circles represent comparision stars.
  The map is shown in Mollweide projection
  in a Galactic coordinate; meridians and parallels 
  are drawn every 45$^{\degr}$.
         }
\end{figure*}

To date, more than 100 interstellar clouds
have been identified at high Galactic latitudes
of $|b|$\,$>$\,20$\degr$, based on
millimeter-wave observations at the frequency
of the J=1-0 transition of CO
(Blitz et al.\ 1984; Magnani et al.\ 1985, 1986, 1996).
Among these HLCs, we searched in
the {\it Mikulski Archive for Space Telescopes}
(MAST) for clouds with high-quality UV data
from the {\it International Ultraviolet Explorer}
(IUE).\footnote{%
  {\sf \url{https://archive.stsci.edu/iue/}}
  }
To this end, we found 32 HLCs with reddened
background O or B stars.
These stars have high quality IUE spectra
obtained with IUE's three cameras:
the Short-Wavelength Prime (SWP) Camera
covering $\simali$1150–-1978$\Angstrom$,
and the two long-wavelength cameras,
i.e., the Long-Wavelength Prime (LWP) 
and the Long-Wavelength Redundant (LWR)
covering $\simali$1850–-3347$\Angstrom$. 
The IUE spectra of these background O and B stars
show prominent interstellar extinction effects,
thus making them ideal sources to derive
the extinction curves for these HLCs.
Figure~\ref{fig:map} shows
the distribution of the selected HLCs
projected on the Galactic longitude–latitude plane.

\begin{figure*}[h]
\centering	
\includegraphics[width=0.7\textwidth]{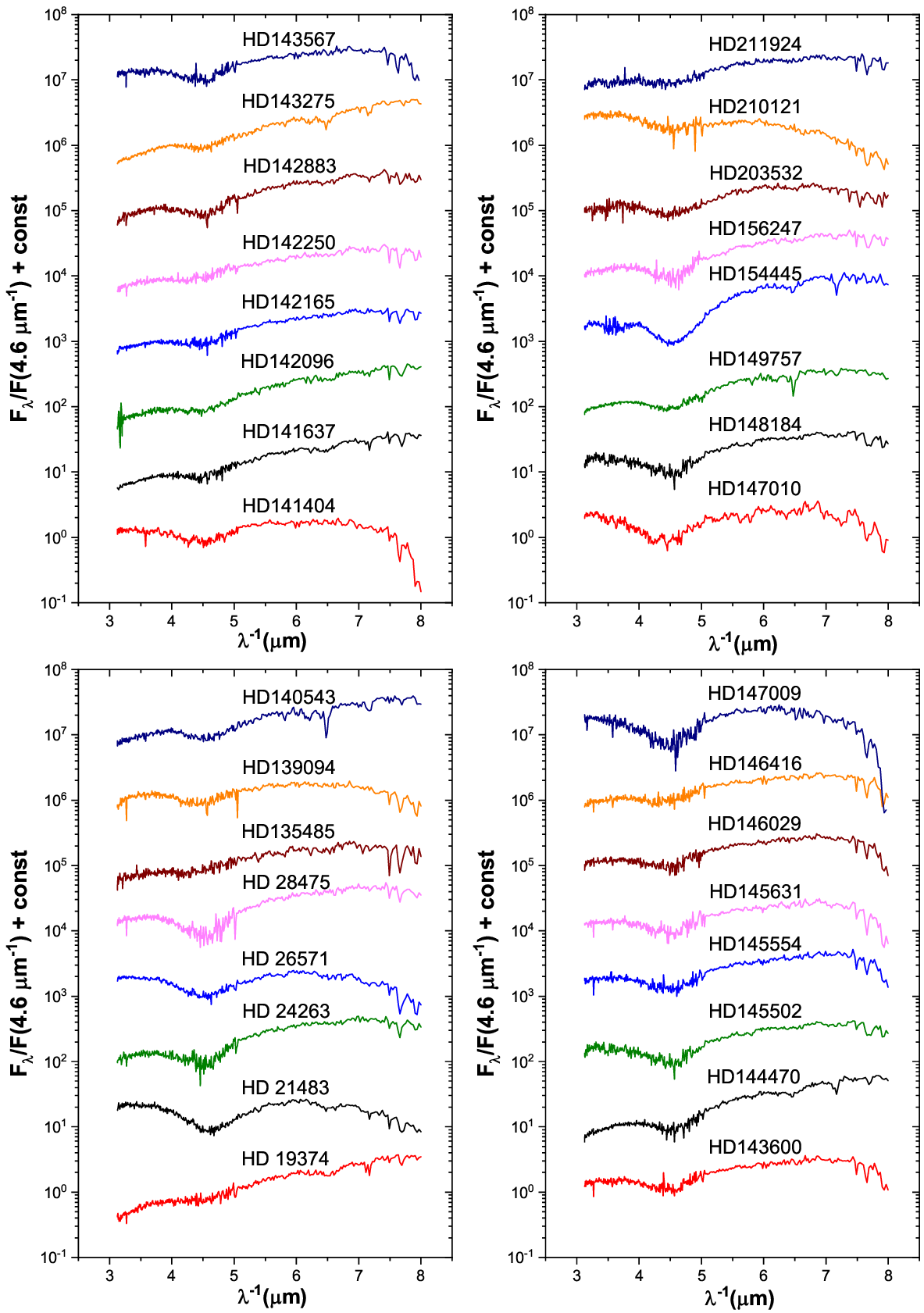}
\vspace{-2mm}
\caption{\footnotesize
        \label{fig:targets}
         Observed IUE spectra of 32 reddened stars.
         The spectra are normalized at
         $\lambda^{-1}=4.6\mum^{-1}$ 
         and vertically separated by 0.5 log space.
         }
\end{figure*}

The spectral data of these reddened, background
stars were processed using the {\it New Spectral
Image Processing System} (NEWSIPS; Nichols \& Linsky 1996)
and then recalibrated using the algorithm described by 
Massa \& Fitzpatrick (2000). If multiple spectra
were present for the SWP, LWP and LWR cameras,
we took an average. The short- and long-wavelength
data were merged at the shortest wavelength of
the SWP camera at 1978$\Angstrom$ to generate
a ``complete'' spectrum spanning a wavelength range 
of $\simali$1200--3000$\Angstrom$.
Finally, the merged short- and long-wavelength
spectra were binned to the instrumental resolution 
of $\simali$5$\Angstrom$.
In Figure~\ref{fig:targets} we show
the observed IUE spectra of these
selected 32 reddened stars.

In addition, we also obtained the UBV magnitudes
of these reddened background stars
from Hiltner et al.\ (1969) and Nicolet (1978) to maintain 
uniformity in the data. Typically, these photometric data
are of high quality, with very small systematic errors
of $\simlt$\,0.01$\magni$. We also utilized the SIMBAD
database to cross-check magnitudes and spectral type
for each reddened background source.
We found that the spectral types given
by SIMBAD are closely consistent with
that adopted here.
Furthermore, we retrieved the near-IR photometry
JHK data for all of our sample sources from the
{\it Two-Micron All-Sky Survey} (2MASS) archive.
To determine the extinction, we used the intrinsic 
colors of FitzGerald (1970) for the UBV photometric bands
and Wegner (1994) for the JHK bands.
Table~\ref{tab:targets} lists the galactic coordinates,
spectral types and UBVJHK photometric magnitudes
of the 32 reddened background stars.
We note that the intrinsic colors adopted here
are close to that of Schmidt-Kaler (1982)
and Pecaut \& Mamajek (2013).

%

The majority of the sources in our sample
have not been observed with
the {\it Space Telescope Imaging Spectrograph}
(STIS) aboard the {\it Hubble Space Telescope} (HST).
While seven of our sources do have available STIS data,
none of these datasets span the wavelength region
around the 2175$\Angstrom$ extinction bump.
Consequently, the available STIS observations
do not provide additional constraints on the strength,
width, or central wavelength of the extinction bump
and the far-UV extinction rise at $\lambda^{-1}>5.9\mum^{-1}$.
As an example, HD\,210121, one of the few targets
with STIS observations, was observed only with
the G430L and G750L gratings, spanning
wavelength ranges of $\simali$2900--5700$\Angstrom$
and $\simali$5240–10,270$\Angstrom$, respectively.
Therefore, the STIS data for HD\,210121 do not extend
into the UV region containing the 2175$\Angstrom$
bump and the far-UV extinction.
For this reason, we rely on the existing IUE datasets
as described above.

\section{Deriving the Extinction Curves}\label{sec:extcurv}
Following Valencic et el.\ (2004) and Gordon et al.\ (2009),
we apply the ``pair method'' to derive the extinction curves
for the sightlines toward background stars
behind the 32 HLCs selected in \S\ref{sec:sample}.
This method compares the photometric or
spectrophotometric data of two stars of
identical spectral types and luminosity classes,
with one star located behind a dust cloud
and  another star, in ideal case, unaffected by dust,
so that there is no obscuration
between the observer and the comparison star
(see Stecher 1965, Massa et al.\ 1983).

\begin{figure*}[h]
\centering	
\includegraphics[width=0.9\textwidth]{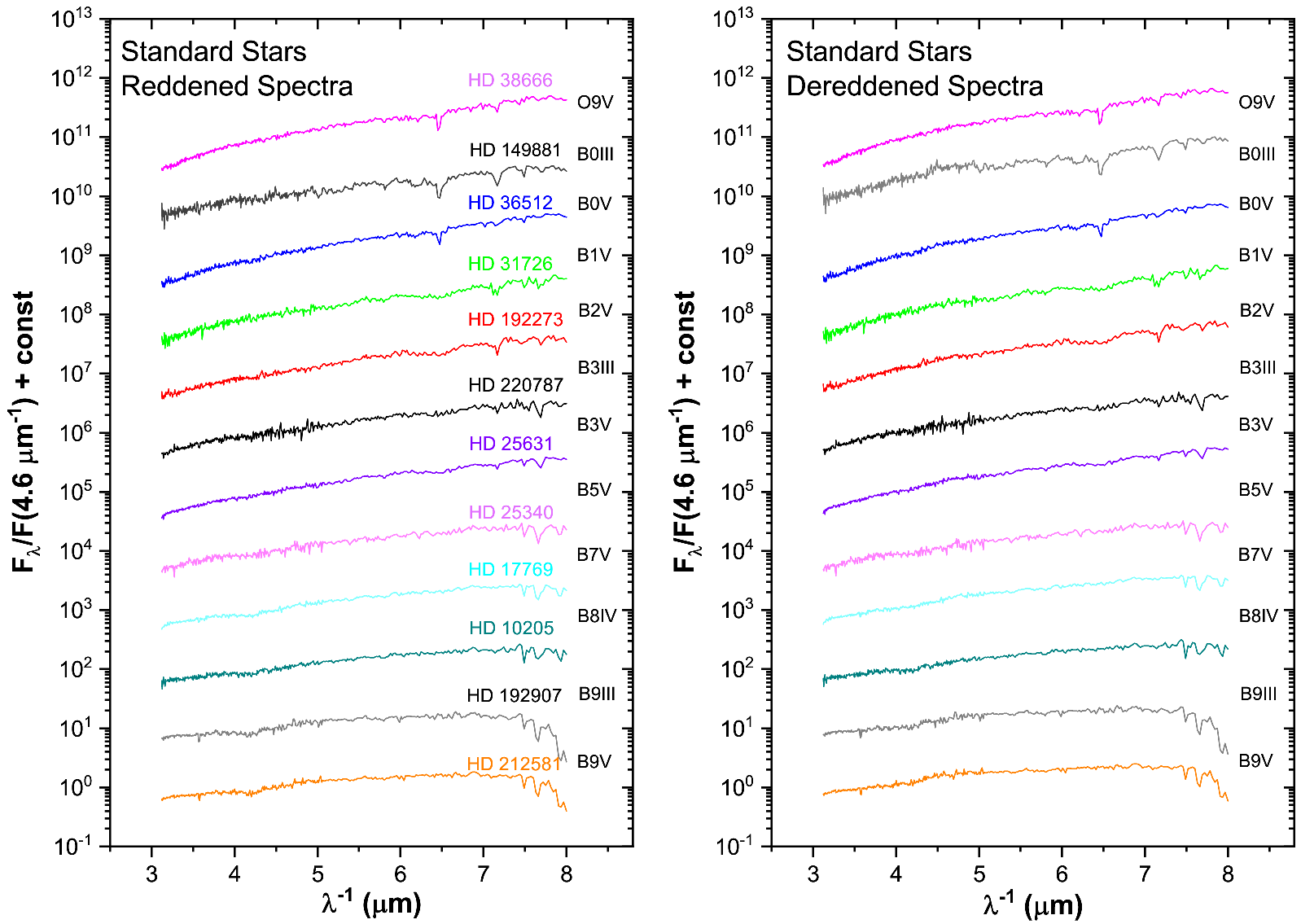}
\vspace{-2mm}
\caption{\footnotesize
        \label{fig:standard_stars}
        Left panel: Observed IUE spectra of comparison stars.
        Right panel: Dereddened spectra of these comparison stars.
        The spectra are normalized at
        $\lambda^{-1}=4.6\mum^{-1}$ 
        and vertically separated by 0.5 log space.
         }
\end{figure*}

Strictly speaking, no sightline in the universe
is free of dust. However, there exist some moderately
reddened stars which can be employed as comparison
stars. To obtain the intrinsic, ``dust-free'' UV spectra
of comparison stars, we select 12 moderately-reddened
stars with $E(B-V)<0.05\magni$ and carefully de-redden
their IUE spectra, in a manner similar to that described
in Cardelli et al.\ (1992) and Gordon et al.\ (2009).
Let $F_\lambda^{\rm c,redd}$ be the original, IUE spectrum
of a comparison star, and $F_\lambda^{\rm c,dere}$ be
its dereddened spectrum. We obtain $F_\lambda^{\rm c,dere}$
as follows:
\begin{equation}
F_\lambda^{\rm c,dere} = F_\lambda^{\rm c,redd}
\cdot\,\exp\left(\frac{A_V}{1.086}
\cdot\frac{A_\lambda}{A_V}\right)
= F_\lambda^{\rm c,redd}
\cdot\,\exp\left\{\frac{R_V\cdot E(B-V)}{1.086}
\left[\frac{1}{R_V}\cdot\frac{E(\lambda-V)}{E(B-V)}
+1\right]\right\} ~~,
\end{equation}
where $A_V$ is the visual extinction,
$A_\lambda$ is the extinction at wavelength $\lambda$,
$R_V\equiv A_V/E(B-V)$ is optical total-to-selective
extinction ratio, and $E(\lambda-V)\equiv A_\lambda-A_V$
is differential extinction (or reddening)
between $\lambda$ and $V$.
We take $R_V=3.1$ for the sightlines toward comparison
stars as they are subject to little extinction and sample
low-density diffuse medium.
We derive the optical reddening $E(B-V)$
by comparing the observed $(B-V)$ colors
with $(B-V)_0$, the intrinsic colors appropriate
for the spectral types and luminosity classes
of the comparison stars
(Hiltner et al.\ 1969, FitzGerald 1970, Nicolet 1978).
%
%
Following Fitzpatrick \& Massa (1990; hereafter FM90),
we parameterize $E(\lambda-V)/E(B-V)$ as follows:
\begin{equation}
E(\lambda-V)/E(B-V) = R_V \left(A_\lambda/A_V - 1\right)
              = c_1 + c_2\,x 
              + c_3\,D(x,\gamma,\xo) 
              + c_4\,F(x) ~~~,
\end{equation}
where $x\equiv\lambda^{-1}$
is the inverse wavelength in $\mu$m$^{-1}$, 
$c_1 + c_2\,x$ represents a linear ``background'',
$D(x,\gamma,\xo)$, a Drude profile of width
$\gamma$ ($\mu$m$^{-1}$) peaking at $\xo$,
describes the 2175$\Angstrom$ extinction bump,
and $F(x)$ represents a far-UV nonlinear rise 
at $\lambda^{-1} > 5.9\mum^{-1}$:
\begin{equation}
D(x,\gamma,\xo) = \frac{x^2}
  {\left(x^2-\xo^2\right)^2 + x^2\gamma^2} ~~~,
\end{equation}
\begin{equation}
F(x) = \left\{\begin{array}{lr} 
0 ~, & x < 5.9\mum^{-1} ~~~,\\

0.5392\,\left(x-5.9\right)^2 
     + 0.05644\,\left(x-5.9\right)^3 ~, 
 & x \ge 5.9\mum^{-1} ~~~.\\
\end{array}\right.
\end{equation}
We apply the Levenberg–Marquardt algorithm
to deredden the IUE spectra of the comparison
stars so that they lack obvious dust extinction
by varying the parameters $c_1$, $c_2$,
$c_3$, $c_4$, $\xo$ and $\gamma$.
In Figure~\ref{fig:standard_stars} we show
the observed and dereddened IUE spectra
of 12 comparison stars.
Table~\ref{tab:deredd} lists
the best-fit parameters $\cpone$, $\cptwo$,
$\cpthree$, $\cpfour$, $\xo$ and $\gamma$,
where
\begin{equation}
c_j^{\prime} = \left\{\begin{array}{lr} 
c_j/R_V + 1 ~, & j=1 ~~~,\\
c_j/R_V ~, & j=2, 3, 4 ~~~.\\
\end{array}\right.
\end{equation}


With the dereddened spectra of the comparison stars derived,
we apply the ``pair-method'' to determine the extinction
curve for each of the 32 target stars, by comparing the IUE
spectrum of a target star with the dereddened spectrum
of a comparison star with nearly identical spectroscopic features. 
This is done by first calculating $E(\lambda-V)$ from
\begin{equation}
E(\lambda-V) = \left(\lambda-V\right)_{\rm targ}
- \left(\lambda-V\right)_{\rm comp} ~~,
\end{equation}
where $\left(\lambda-V\right)_{\rm targ}$ 
and $\left(\lambda-V\right)_{\rm comp}$
are respectively the colors
of the target and comparison stars
with respect to the $V$ band,
and then calculating $A_\lambda/A_V$ from
\begin{equation}
A_\lambda/A_V = E(\lambda-V)/A_V +1 ~~.
\end{equation}
Following Gordon et al.\ (2009),
the visual extinction $A_V$ is obtained
by utilizing the IR extinction curve of
Rieke et al.\ (1989) and extrapolating
the JHK $E(\lambda-V)$ curves
to infinite wavelength.
The resulting extinction curves
at $3.3 <\lambda^{-1} < 8.7\mum^{-1}$
for all 32 HLC sources are show in
Figures~\ref{fig:extcurv1}--\ref{fig:extcurv4}.
Following Gordon et al.\ (2009),
we estimate the uncertainties on
$A_\lambda/A_V$ as follows:
\begin{equation}
\bigsigma^2\left(A_\lambda/A_V\right) =  
\left(\frac{A_\lambda}{A_V}\right)^2
\left\{ \frac{\bigsigma\left[E(\lambda-V)\right]}
{E(\lambda-V)}\right\}^2
+ \left(\frac{A_\lambda}{A_V}\right)^2
\left\{ \frac{\bigsigma(A_V)}{A_V}\right\}^2
+ \bigsigma_{\rm comp}^2(\lambda) ~~,
\end{equation}  
where $\bigsigma\left[E(\lambda-V)\right]$
is the uncertainty on $E(\lambda-V)$,
and $\bigsigma_{\rm comp}(\lambda)$ 
is the uncertainty due to dereddening
the comparison stars.


\section{Discussion}\label{sec:discussion}
\begin{figure*}
\centering	
\vspace{-5mm}
\includegraphics[width=0.9\textwidth]{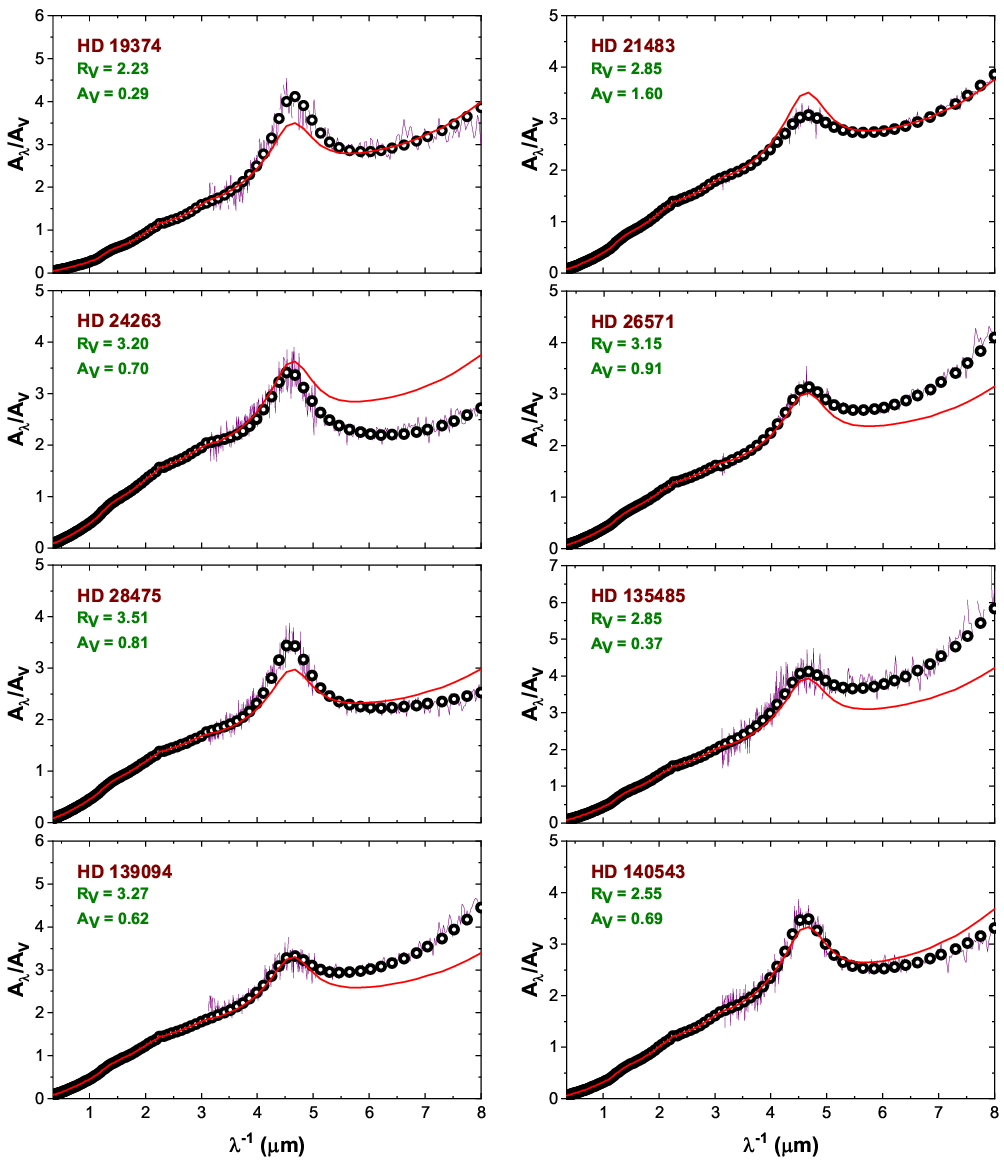}
\vspace{-2mm}
\caption{\footnotesize
         \label{fig:extcurv1}
Extinction curves of the high Galactic latitude sightlines
toward HD\,19374, HD\,21483, HD\,24263, HD\,26571,
HD\,28475, HD\,135485, HD\,139094, and HD\,140543.
The extinction curves derived from
the IUE spectra are shown as gray lines.
The open black circles show 
the FM90 parametrization at
$3.3 <\lambda^{-1} < 8.7\mum^{-1}$ and 
the CCM parametrization at
$\lambda^{-1} < 3.3\mum^{-1}$.
The solid red line plots the CCM parametrization 
at $0.35<\lambda^{-1}<8.7\mum^{-1}$.         
         }
\end{figure*}

\begin{figure*}
\centering	
\vspace{-5mm}
\includegraphics[width=0.9\textwidth]{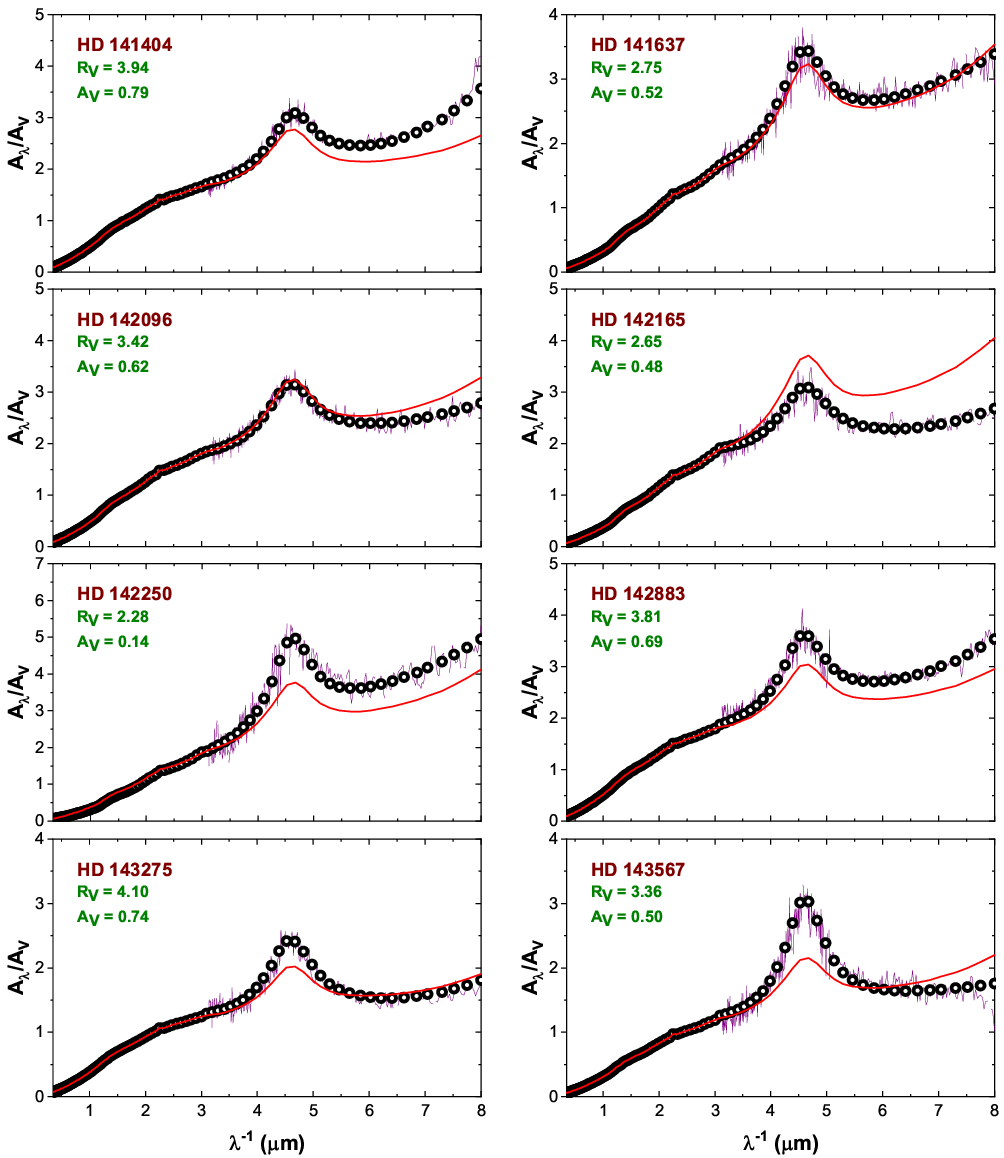}
\vspace{-2mm}
\caption{\footnotesize
        \label{fig:extcurv2}
        Same as Figure~\ref{fig:extcurv1}
        but for HD\,141404, HD\,141637, HD\,142096, HD\,142165, 
        HD\,142250, HD\,142883, HD\,143275, and HD\,143567.
         }
\end{figure*}

\begin{figure*}
\centering	
\vspace{-5mm}
\includegraphics[width=0.9\textwidth]{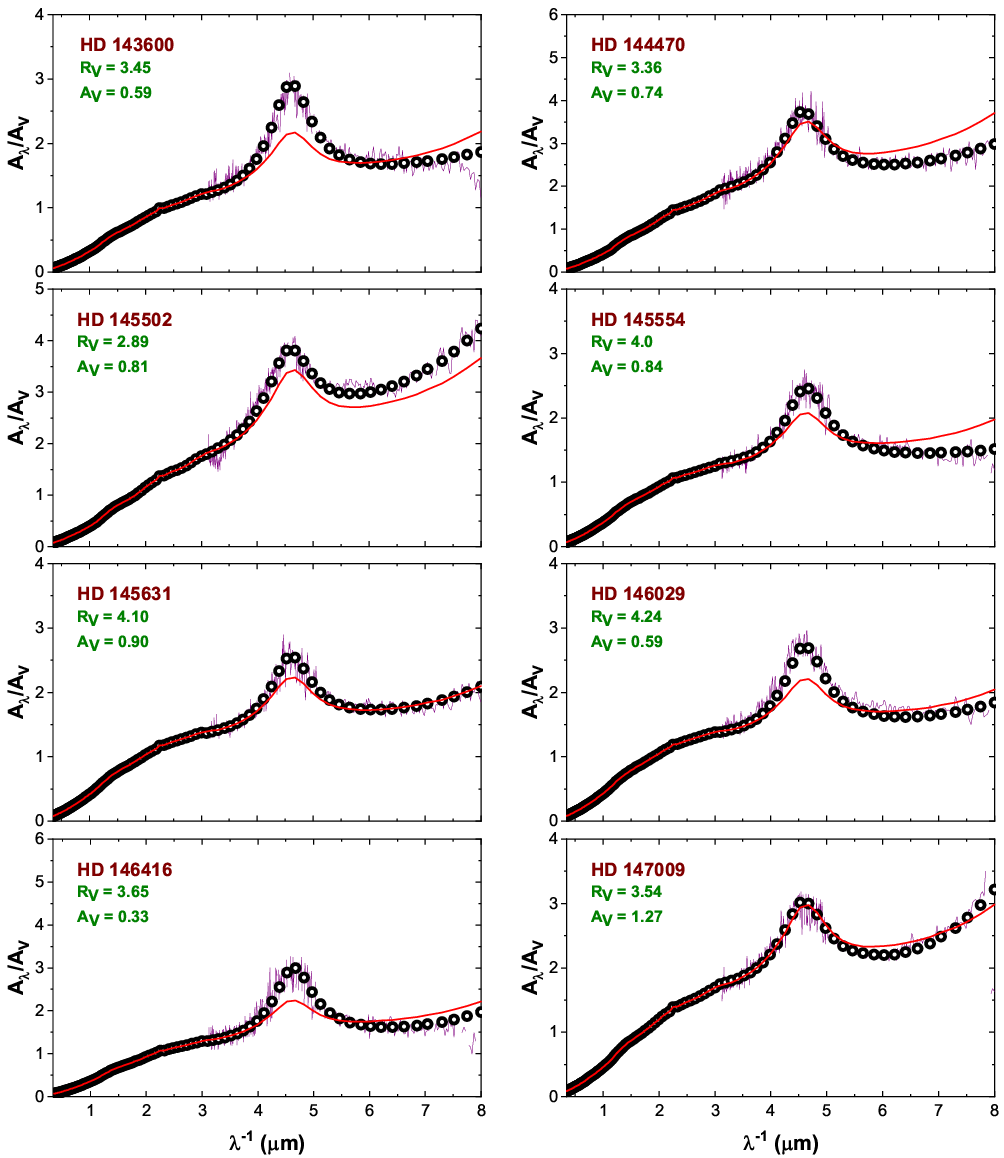}
\vspace{-2mm}
\caption{\footnotesize
        \label{fig:extcurv3}
        Same as Figure~\ref{fig:extcurv1}
        but for HD\,143600, HD\,144470, HD\,145502. HD\,145554, 
        HD\,145631, HD\,146029, HD\,146416, and HD\,147009.
         }
\end{figure*}

\begin{figure*}
\centering	
\vspace{-5mm}
\includegraphics[width=0.95\textwidth]{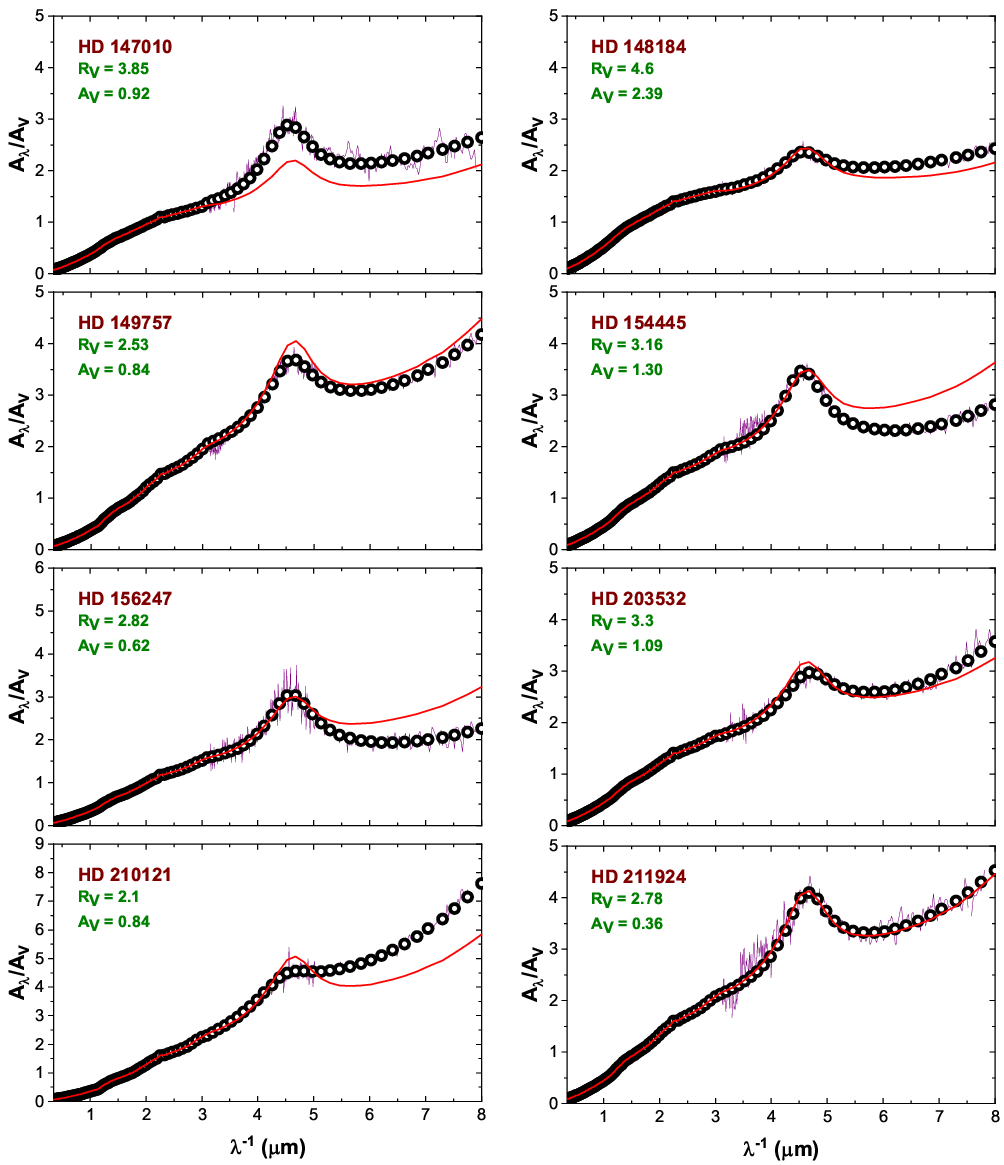}
\vspace{-2mm}
\caption{\footnotesize
  \label{fig:extcurv4}
  Same as Figure~\ref{fig:extcurv1}
  but for HD\,147010, HD\,148184,
  HD\,149757, HD\,154445, HD\,156247,
  HD\,203532, HD\,210121, and HD\,211924.
         }
\end{figure*}

Cardelli, Clayton, \& Mathis (1989; CCM) found that
the mean interstellar extinction curves of both diffuse
and dense regions of the interstellar medium,
from the near-IR through the optical and UV,
can be well represented by a mean relationship
which only depends upon the single parameter $R_V$.
To compare the extinction curves derived
from the IUE data for the 32 HLC sightlines
with the CCM curves, we also show
in Figures~\ref{fig:extcurv1}--\ref{fig:extcurv4}
the CCM extinction curve for each sightline
calculated from the orresponding $R_V$ value.
Among the 32 HLC sightlines, the extinction curves
of several sightlines can be closely represented
by the CCM curves, including
HD\,141637 (see Figure~\ref{fig:extcurv2}),
HD\,147009 (see Figure~\ref{fig:extcurv3}),
HD\,203532 (see Figure~\ref{fig:extcurv4}), and
HD\,211924 (see Figure~\ref{fig:extcurv4}).
A number of sightlines exhibit a far-UV
extinction rise considerably steeper than
the CCM prediction
(e.g., HD\,26571, HD\,135485, HD\,139094,
HD\,141404, and HD\,210121).
In contrast, the far-UV extinction rises of
several other sightlines are substantially
flatter than the CCM prediction
(e.g., HD\,24263, HD\,142165, HD\,144470,
HD\,154445, and HD\,156247).
Also, several sightlines display an extinction
bump at 2175$\Angstrom$ appreciably narrower
than the CCM prediction
(e.g., HD\,28475, HD\,143567, HD\,143600,
HD\,146029, and HD\,146416).
On the other hand, the 2175$\Angstrom$
extinction bumps of the sightlines toward
 HD\,210121 and HD\,21483 are appreciably
broader and weaker than the CCM prediction.
These imply that the size distrubutions of
the dust grains in HLCs differ considerably
from one cloud to another
and from that of the Galactic diffuse ISM.

While the CCM paramterization describes well
the average behavior of the Galactic extinction curves,
the FM90 paramterization is capable of fitting
essentially all individual UV extinction curves of
the Milky Way and the Large and Small Magellanic Clouds.
We adopt a modified version of the FM90 paramterization
to characterize the UV extinction curves
at $3.3 <\lambda^{-1} < 8.7\mum^{-1}$
derived in \S\ref{sec:extcurv} for the HLC sightlines:
\begin{equation}\label{eq:A2AV}
A_\lambda/A_V = c_1^{\prime} + c_2^{\prime}\,x 
              + c_3^{\prime}\,D(x,\gamma,\xo) 
              + c_4^{\prime}\,F(x) ~~~.
\end{equation}
The modified FM90 parameters $c_j^{\prime}$
($j$\,=\,1, 2, 3, and 4) as well as $\xo$ and $\gamma$
were determined by fitting the analytical formula
(see eq.\,\ref{eq:A2AV}) to an observed extinction
curves. The fitting procedure employed
the Levenberg–Marquardt nonlinear
least-square minimization algorithm
to obtain the set of parameters
($c_j^{\prime}$, $\xo$, and $\gamma$) 
that best reproduce the observed extinction curve,
while accounting for observational uncertainties.
Following Gordon et al.\ (2003), the parameters were
derived using a three-step fitting procedure
over the wavelength range of 1250–-2700$\Angstrom$.
Note that the FM90 parameterization is not considered
reliable at wavelengths longer than 2700$\Angstrom$,
while the spectra were truncated at 1250$\Angstrom$
to avoid contamination from the interstellar HI Ly$\alpha$
absorption feature at 1215$\Angstrom$.
The 2175$\Angstrom$ bump is represented
by a Drude profile of width $\gamma$
and peak $\xo$, which is the theoretical profile 
expected for a classic damped harmonic oscillator
(see Li 2009), whereas the far-UV rise is described
by a nonlinear curvature term that becomes significant
at $x\gtsim5.9\mum^{-1}$. The resulting best-fit
parameters provide a quantitative description of
the extinction properties along each sightline and
enable direct comparisons of grain characteristics
and extinction behavior across different interstellar
environments.
%
%
To be complete, we have also employed
the $R_V$-based CCM parameterization
to represent the extinction
at $1.1 < \lambda^{-1} < 3.3\mum^{-1}$,
where $R_V$ is calculated from $A_V$
and $E(B-V)$.
We tabulate in Table~\ref{tab:extpara}
the FM90 parameters $\cpone$, $\cptwo$,
$\cpthree$, $\cpfour$, $\xo$ and $\gamma$,
and show in Figures~\ref{fig:extcurv1}--\ref{fig:extcurv4}
the FM90 curves. It is apparent that the FM90
paramterization indeed closely fits the UV extinction
curves of all the HLC sightlines.

%

\begin{figure*}[htp]
\begin{center}
\includegraphics[width=0.4\textwidth]{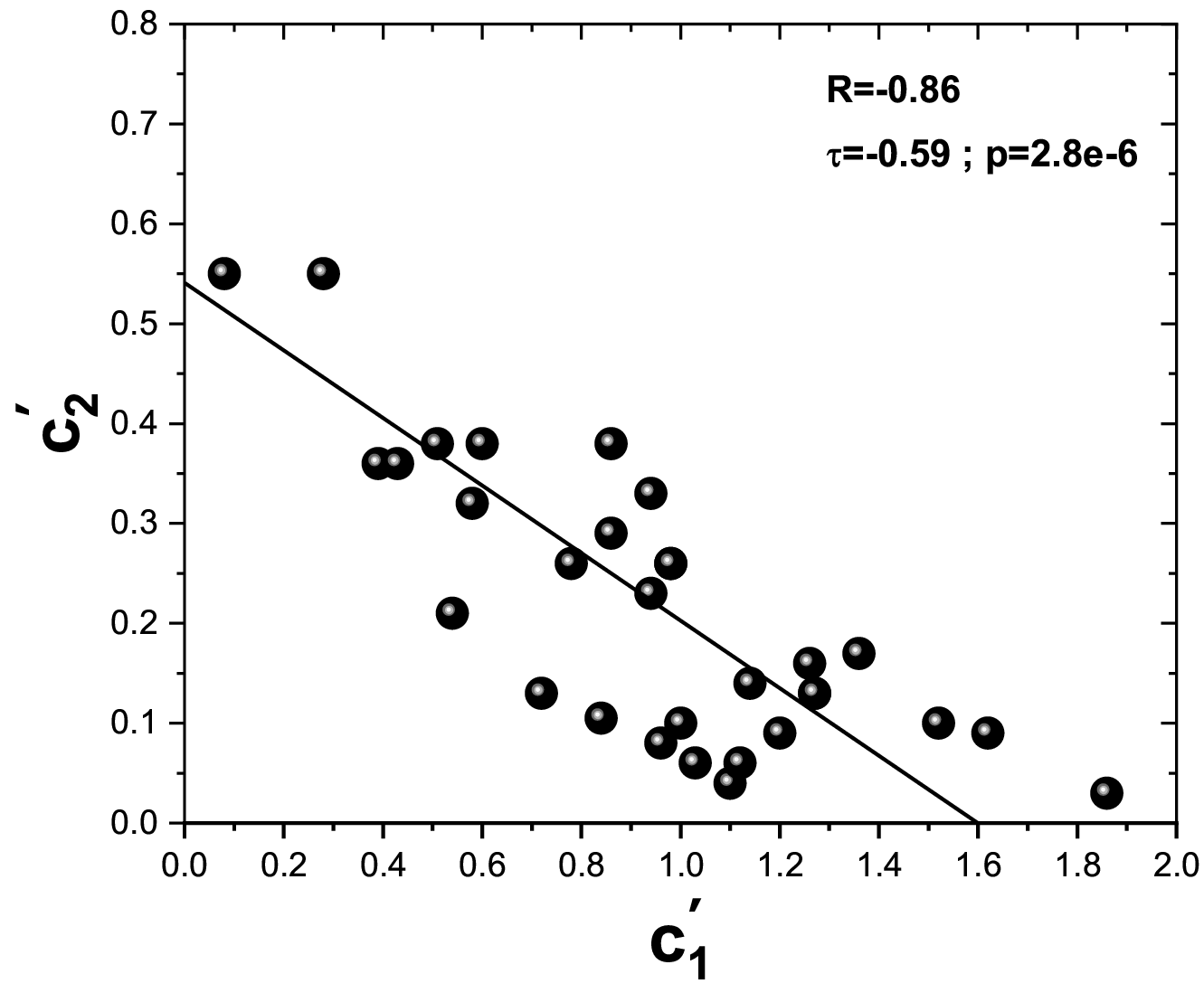}
\includegraphics[width=0.4\textwidth]{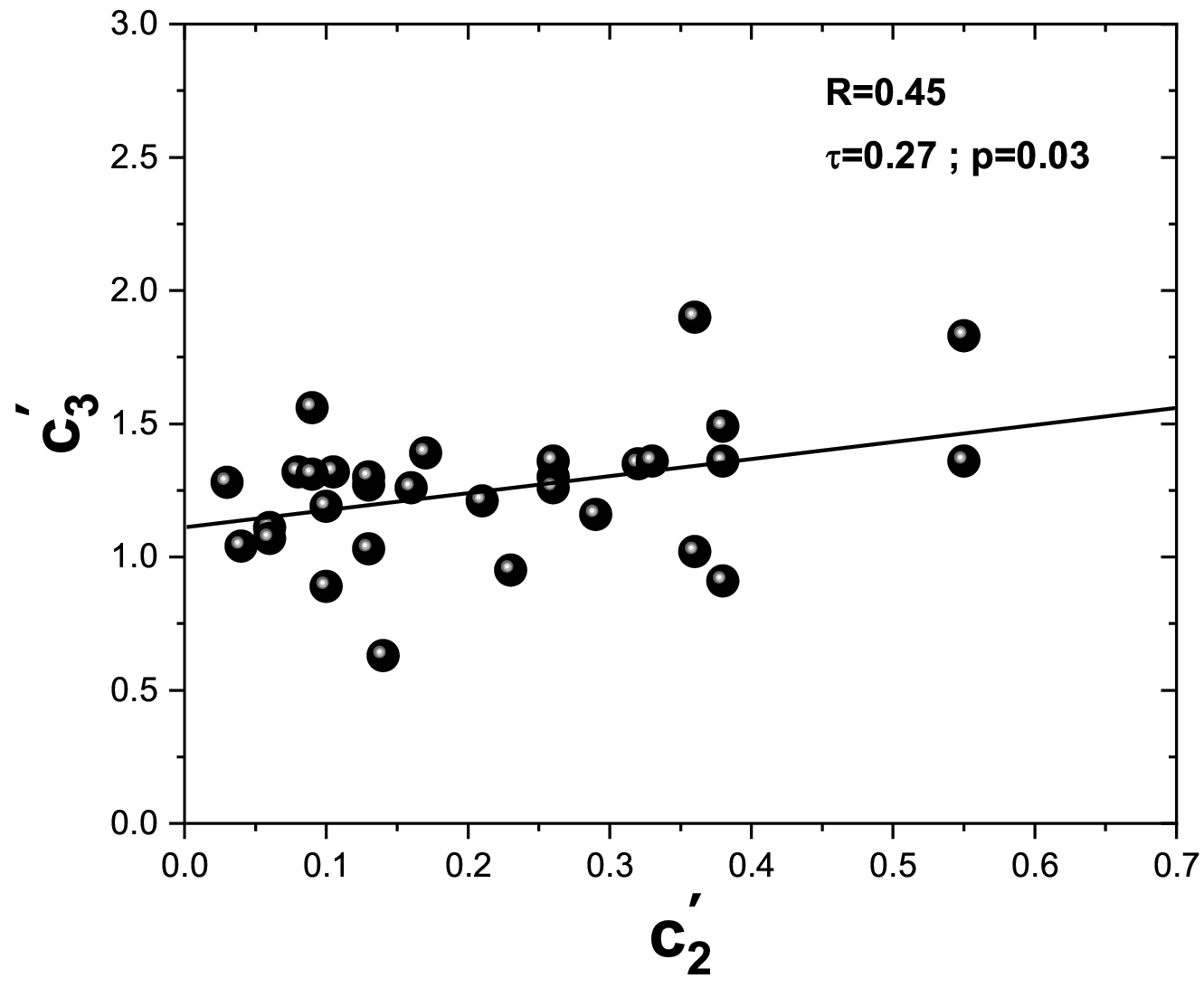}\\
\includegraphics[width=0.4\textwidth]{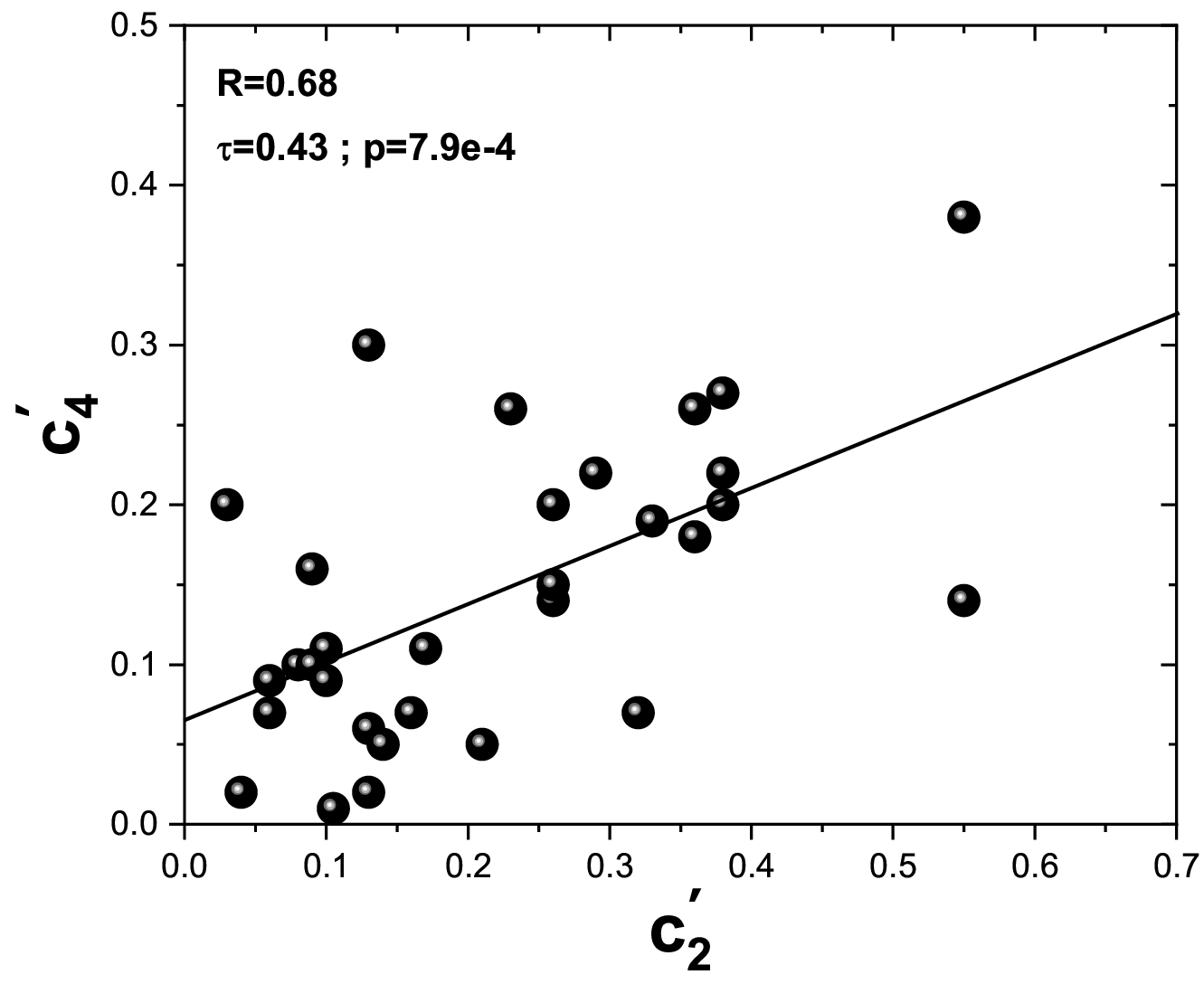}
\includegraphics[width=0.4\textwidth]{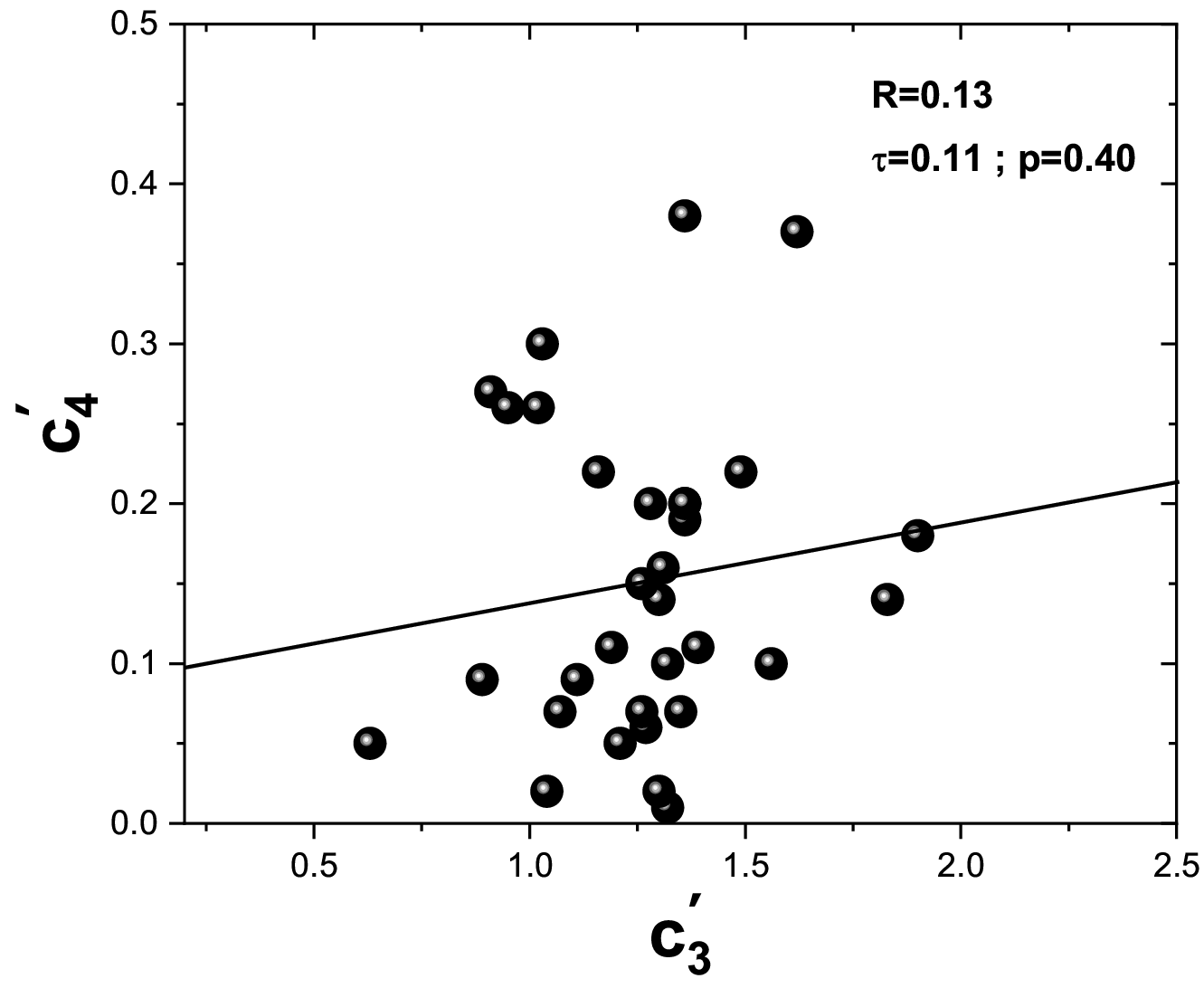}\\
\end{center}
\vspace{-5mm}
\caption{\label{fig:cpara} 
Correlations between the modified FM90 parameters:
$c_1^{\prime}$ vs.  $c_2^{\prime}$ (upper left),   
$c_2^{\prime}$ vs.  $c_3^{\prime}$ (upper right),
$c_2^{\prime}$ vs.  $c_4^{\prime}$ (bottom left), and
$c_3^{\prime}$ vs.  $c_4^{\prime}$ (bottom right).
Also labelled are the Pearson correlation
coefficient $R$ and the Kendall's $\tau$
coefficient and the significance level $p$.
Note the anti-correlation between
the intercept $c_1^{\prime}$ and
the slope $c_2^{\prime}$ of the linear
``background'' extinction, and the non-correlation
between the 2175$\Angstrom$ extinction bump
$c_3^{\prime}$ and the non-linear extinction rise
$c_4^{\prime}$. 
}
\vspace{-3mm}
\end{figure*}

We examine the correlations among
the $c_j^{\prime}$ parameters and
show the correlation results
in Figure~\ref{fig:cpara}.
Apparently, there is a clear anti-correlation
between $c_1^{\prime}$---the intercept
of the linear extinction background---and
$c_2^{\prime}$---the slope of the linear extinction
background. Such an anti-correlation between
$c_1^{\prime}$ and $c_2^{\prime}$ has been noted
before (Fitzpatrick \& Massa 1988,
Jenniskens \& Greenberg 1993,
Valencic et al.\ 2004),
suggesting that $c_1^{\prime}$ and $c_2^{\prime}$
are not independent.
On the other hand,
there is no correlation between 
$c_2^{\prime}$ and $c_3^{\prime}$,
consistent with earlier studies of
Fitzpatrick \& Massa (1988),
Jenniskens \& Greenberg (1993),
and Valencic et al.\ (2004).
This implies that the linear background extinction
and the 2175$\Angstrom$ extinction bump are
well defined, despite that $c_1^{\prime}$
and $c_2^{\prime}$ are not independent.
Differing from previous studies
(e.g., see Fitzpatrick \& Massa 1988),
$c_2^{\prime}$ appears to correlate
with $c_4^{\prime}$.
However, we should note that 
the scatter is rather large
for the 32 sightlines studied here. 
Finally, Figure~\ref{fig:cpara} shows
that $c_3^{\prime}$ and $c_4^{\prime}$
are not correlated, suggesting that the bump
and the far-UV extinction arise from different
carriers, as originally argued by
Greenberg \& Chlewicki (1983).
Nevertheless, we note that, while the FM90
parametrization provides an excellent
mathematical description of the UV extinction
at $\lambda^{-1}>3.3\mum^{-1}$,
the distinction between the linear rise 
(measured by $c_1$ and $c_2$)
and the FUV non-linear rise (measured by $c_4$)
probably has little physical significance 
since there is no substance known that shows 
the corresponding extinction of any of them.
Xiang et al.\ (2017) proposed to decompose
the extinction curve into three Drude-like functions
composed of  the visible/near-IR component, 
the 2175$\Angstrom$ bump, 
and the far-UV extinction 
at $\lambda^{-1}>5.9\mum^{-1}$.
They argued that the wavelength-integrated 
extinction derived from this decomposition technique 
best measures the strength of the extinction.
As dust grains scatter and absorb light
most effectively at wavelengths comparable
to their sizes (see Li 2009), the far-UV extinction
provides {\it qualitative} information on the dust sizes,
that is, a population of grains smaller than
$\simali$20\,nm are responsible for the far-UV
extinction. However, the far-UV extinction cannot
tell the exact size distribution of these grains
(see eq.\,1 of Wang et al.\ 2015).
In contrast, near- and mid-IR observations
of the thermal emission from these grains
which undergo stochastic heating by single
starlight photons at high Galactic latitudes
(Draine \& Li 2001) will enable us to
constrain their size distributions
(e.g., see Li \& Draine 2001).
In this aspect, the {\it James Webb Space Telescope}
will be instrumental in detecting the near-
and mid-IR emission of HLCs.

\begin{figure*}[htp]
\begin{center}
\includegraphics[width=0.8\textwidth]{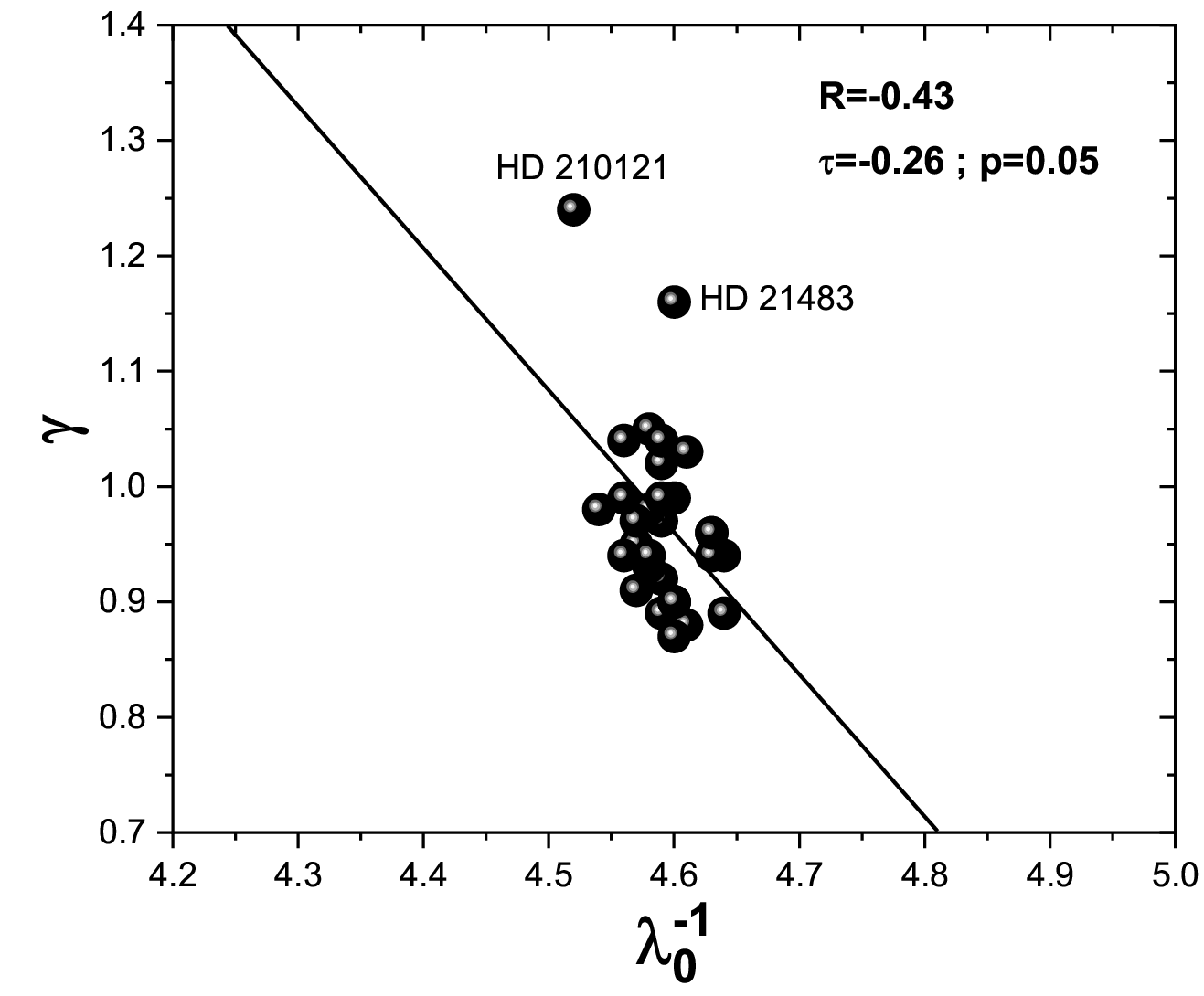}
\end{center}
\vspace{-5mm}
\caption{\label{fig:xo_vs_gamma}
Variations of the the peak positions ($\xo$)
and widths ($\gamma$) of the 2175$\Angstrom$
extinction bump derived for the 32 high Galactic
latitude sightlines. 
Also labelled are the Pearson correlation
coefficient $R$ and the Kendall's $\tau$
coefficient and the significance level $p$.
Note the invariant bump position
and variable bump width.
}
\vspace{-3mm}
\end{figure*}

The most striking characteristics of
the 2175$\Angstrom$ extinction bump
are the invariant central
wavelength and variable width:
its peak position at 2175$\Angstrom$
is remarkably constant, while its width
varies from one line of sight to another
(see Whitett 2022, Wang et al.\ 2023, 2026
and references therein).
This is also true for the 32 high Galactic
latitude sightlines studied here.
As shown in Figure~\ref{fig:xo_vs_gamma},
the peak position of the bump ($\xo$)
is essentially invariant while the bump
width ($\gamma$) varies
from $\simali$0.87 (HD\,143567)
to $\simali$1.16$\mum^{-1}$ (HD\,210121) 
and $\simali$1.26$\mum^{-1}$ (HD\,21483),
with a mean bump width of $\simali$0.95$\mum^{-1}$.
This is consistent with Valencic et al.\ (2004)
who derived an average bump width
of $\simali$$0.92\pm0.12\mum^{-1}$
for 417 Galactic sightlines.
While the exact carrier of the 2175$\Angstrom$ bump
remains unidentified (see Draine 1989)
and a thorough discussion is out of
the scope of this work, we note that
polycyclic aromatic hydrocarbon (PAH) molecules
(Joblin et al.\ 1992, Li \& Draine 2001,
Steglich et al.\ 2013, Lin et al.\ 2023, 2025),
carbon nanoparticles containing
turbostratic graphitic basic structural
units (J\"ager et al.\ 2008, Schnaiter et al.\ 1998)
or tetrahedron carbon (Sheng et al.\ 2011, Ma et al.\ 2020),
and carbon buckyonions
composed of spherical concentric fullerene shells
(Chhowalla et al.\ 2003,  Iglesias-Groth et al.\ 2003,
Ruiz et al.\ 2005, Li et al.\ 2008)
are among the promising candidate materials.

\begin{figure*}[h]
\centering	
\vspace{-5mm}
\includegraphics[width=1.0\textwidth]{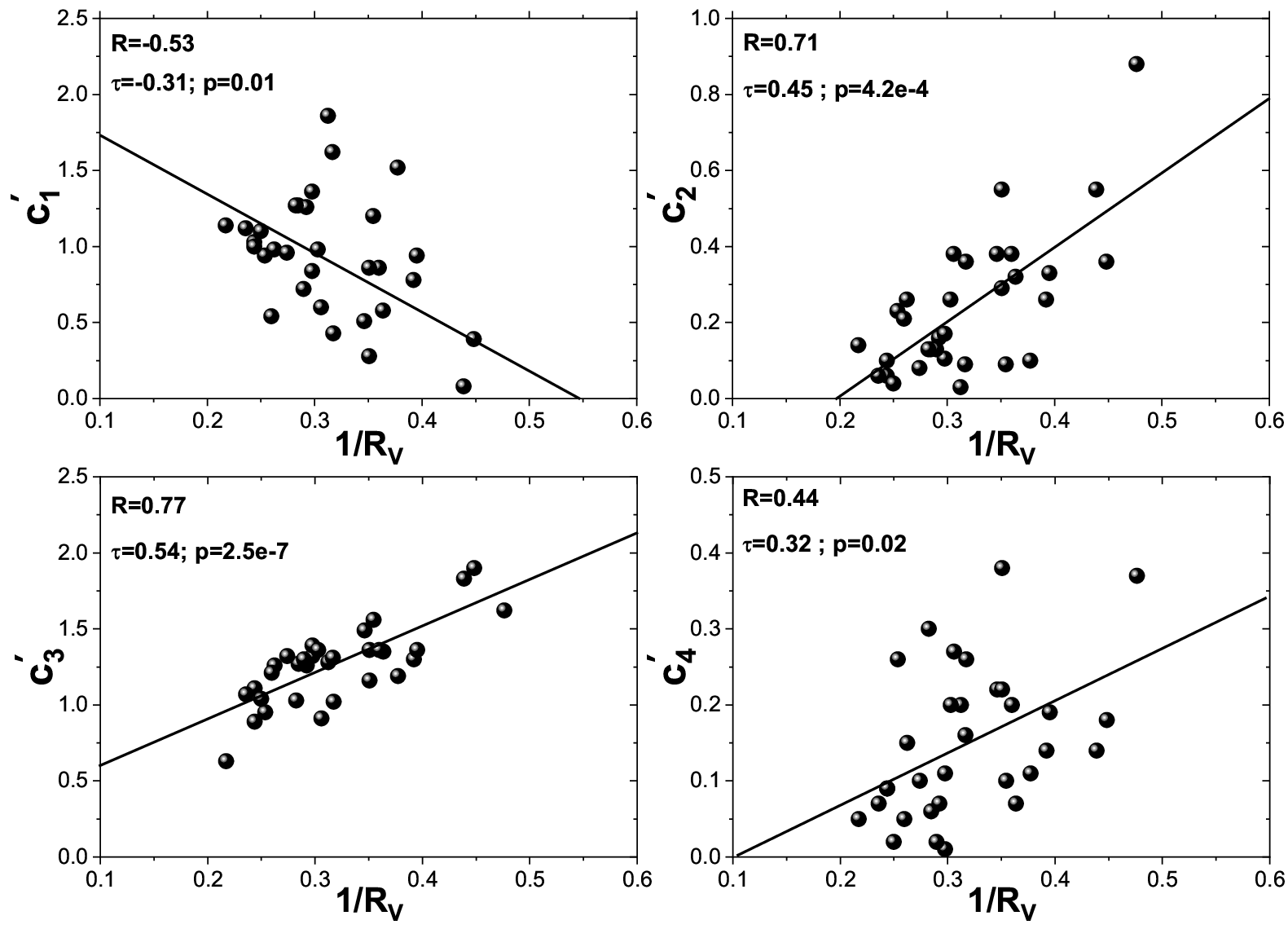}
\vspace{-2mm}
\caption{\footnotesize
        \label{fig:cjp_vs_RV}
        Dependencies of the FM90 parameters
        $\cpone$, $\cptwo$, $\cpthree$,
        and $\cpfour$ on $R_V^{-1}$.
        Also labelled are the Pearson correlation
        coefficient $R$ and the Kendall's $\tau$
        coefficient and the significance level $p$.
         }
\end{figure*}

\begin{figure*}[h]
\centering	
\vspace{-5mm}
\includegraphics[width=1.0\textwidth]{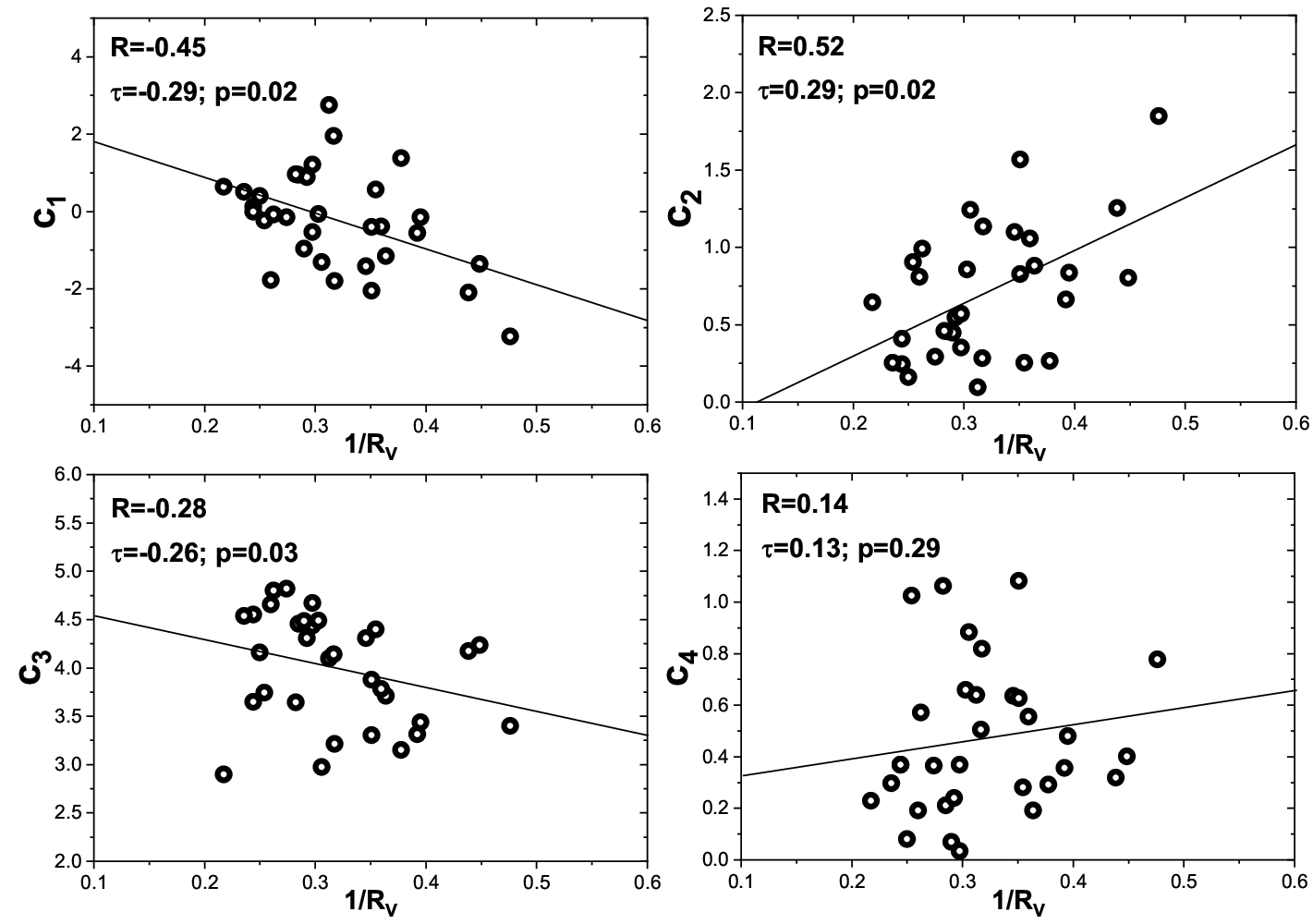}
\vspace{-2mm}
\caption{\footnotesize
        \label{fig:cj_vs_RV}
         Dependencies of the FM90 parameters
        $\cone$, $\ctwo$, $\cthree$,
        and $\cfour$ on $R_V^{-1}$.
        Also labelled are the Pearson correlation
        coefficient $R$ and the Kendall's $\tau$
        coefficient and the significance level $p$.
         }
\end{figure*}

We explore the relationships between the FM90
parameters $c_j^{\prime}$ ($j$\,=\,1, 2, 3 and 4)  and $R_V$.
As shown in Figure~\ref{fig:cjp_vs_RV},
with a Pearson correlation coefficient 
of $R\approx-0.53$
and a Kendall $\tau\approx-0.31$ 
and $p\approx0.01$, $\cpone$, the intercept
of the linear component of the extinction,
somewhat anti-correlates with $R_V^{-1}$.
In contrast, the slope of the linear component $\cptwo$,
the strength of the 2175$\Angstrom$ bump $\cpthree$,
and the strength of the nonlinear component of
the far-UV extinction $\cpfour$,
all appear to correlate with $R_V^{-1}$.
Such relations have previously been
reported by Valencic et al.\ (2004).
In view of the fact that $c_j^{\prime}=c_j/R_V$
where $j$\,=\,2, 3 and 4, the correlation of
$\cptwo$, $\cpthree$, and $\cpfour$
with $R_V^{-1}$ may merely reflect
the proportionality between $c_j^{\prime}$
and $c_j$. To this end, we examine
in Figure~\ref{fig:cj_vs_RV}
the relationships between the FM90 parameters
$c_j$ ($j$\,=\,1, 2, 3 and 4) and $R_V^{-1}$.
Figure~\ref{fig:cj_vs_RV} shows that,
unlike $\cpthree$ and $\cpfour$,
$\cthree$ and $\cfour$ are not correlated
with $R_V^{-1}$,  while the $\cone$--$R_V^{-1}$
anti-correlation and the $\ctwo$--$R_V^{-1}$
correlation persist, but to a lesser degree.
The correlation between $\ctwo$ and $R_V^{-1}$
is consistent with the scenario that, in sightlines
of smaller $R_V$ values, the far-UV extinction
rise is steeper. The anti-correlation between
$\cone$ and $R_V^{-1}$ merely reflects 
the anti-correlation between
$\cone$ and $\ctwo$
(see Figure~\ref{fig:cpara}; also see Figure~4
of Fitzpatrick \& Massa 1988, and Figure~2
of Valencic et al.\ 2004).
Nevertheless, as mentioned earlier,
the linear extinction component is largely
a mathematical decompositional outcome
and no dust components produce
such a linear extinction,
$\cone$ and $\ctwo$ are more mathematical
than physical. Therefore, we shall not over-interpret
the dependencies of $\cone$ and $\ctwo$ on $R_V^{-1}$.

\begin{figure*}[h]
\centering	
\vspace{-5mm}
\includegraphics[width=0.95\textwidth]{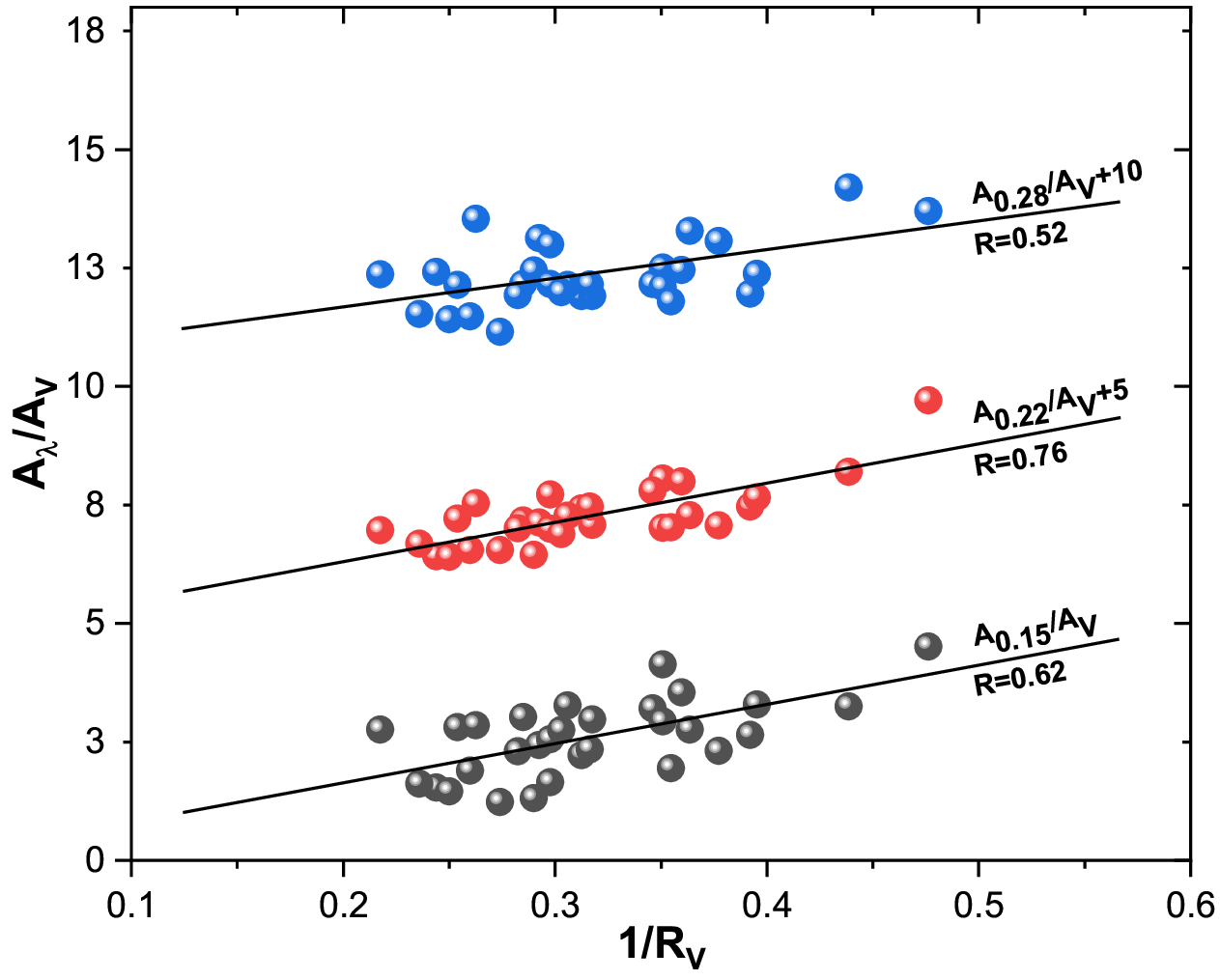}
\vspace{-2mm}
\caption{\footnotesize
        \label{fig:Alambda_RV}
         The extinction ratio $A_\lambda/A_V$ 
         plotted against $R_V^{-1}$ 
         at selected wavelengths.
         The subscripts refer to the wavelength
         (e.g., $A_{0.15}$ refers to the extinction
         at $\lambda=0.15\mum$).
         The data for $\lambda=0.15, 0.22, 0.28\mum$
         have been shifted vertically by the amount
         indicated in order to separate them.
         } 
\end{figure*}

Figure~\ref{fig:Alambda_RV} examines
the variations of $A_\lambda/A_V$ with $R_V^{-1}$
for $\lambda$\,=\,0.15$\mum$, 0.22$\mum$, 
and 0.28$\mum$. It is apparent that 
there is a clear linear relationship 
between $A_\lambda/A_V$ and $R_V^{-1}$
at all wavelengths, confirming the earlier findings
of Cardelli et al.\ (1989).
The correlation between $R_V^{-1}$ 
and the extinction at different wavelengths
suggests the existence of 
a common process that simultaneously 
modifies all parts of the extinction curve.
As the extinction at 
a particular wavelength $\lambda$
is dominated by grains of 
a particular size $a\sim\lambda/2\pi$
(see Li 2009), the fact that $A_\lambda/A_V$
correlates with $R_V^{-1}$ at different wavelengths
implies that the process which produces changes
in extinction must operate effectively and rather
continuously over a wide range of grain sizes.

\begin{figure*}[h]
\centering	
\vspace{-15mm}
\includegraphics[width=0.6\textwidth]{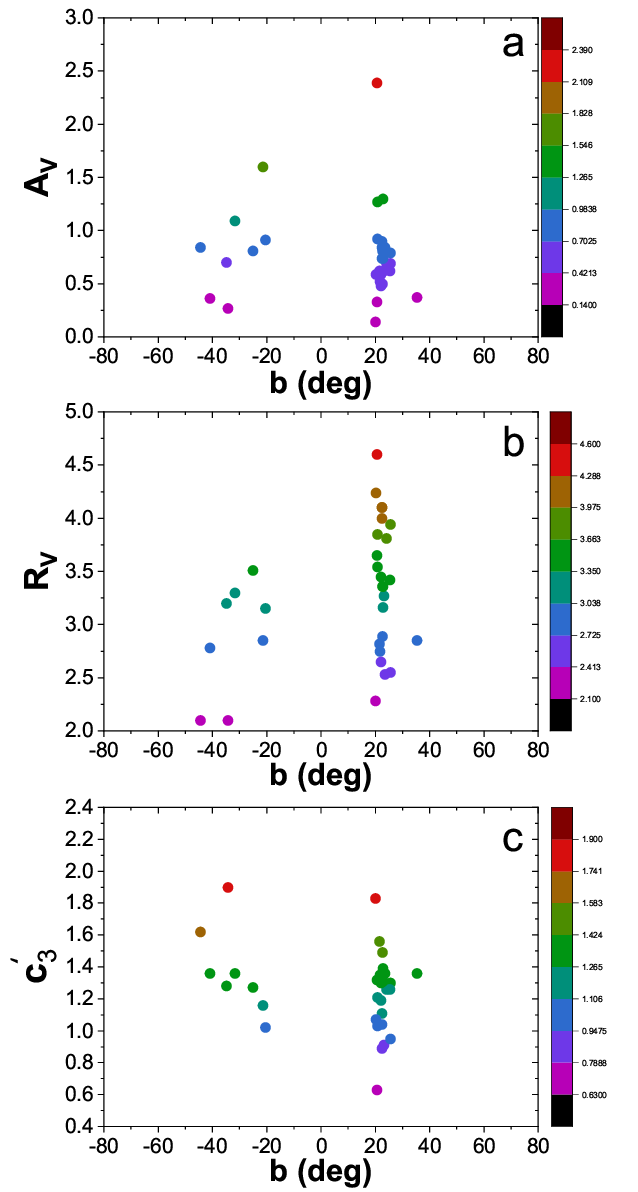}
\vspace{-3mm}
\caption{\footnotesize
        \label{fig:spatial_variation}
        Spatial variation of $A_V$, $R_V$,
        and the 2175$\Angstrom$ extinction
        bump strength $\cpthree$
        with the Galactic latitude $b$.
         }
\end{figure*}

Finally, we examine in Figure~\ref{fig:spatial_variation}
the spatial variations of $A_V$, $R_V$
and the 2175$\Angstrom$ extinction bump strength
$\cpthree$ with the Galactic latitude $b$.
Among our 32 HLC sightlines, only the sightline
toward HD\,148184 has $A_V>2\magni$.
Most of the sightlines (27/32) have $A_V<1\magni$,
confirming the diffuse or translucent nature of HLCs.
The HLCs span a wide range of $R_V$ values,
from $\simali$2.0 to $\simali$4.6.
Figure~\ref{fig:spatial_variation} clearly shows
that neither $A_V$, $R_V$ nor $\cpthree$ exhibits
any systematic variations with latitudes,
implying that the dust size and compositional
properties do not systematically vary with latitudes.

\section{Summary\label{sec:summary}}
We have selected 32 sightlines toward reddened
background stars at high Galactic latitudes
with $|b|$\,$>$\,20$\degr$
for which high-quality IUE spectra are available.
The ``pair-method'' is employed to derive
the UV interstellar extinction curves
for these sightlines. It is found that the extinction
properties of these sightlines exhibit no systematic
variations with the Galactic latitudes, although they
do show appreciable sightline-to-sightline variations.

\acknowledgments
We thank B.T.~Draine, Q.~Li, Q.~Lin, Q.~Wang
and X.J.~Yang for helpful discussions.
We also thank J.L.~Linsky for his comments
and suggestions that substantially improved
the quality and presentation of this paper.
We are supported in part by
NSF grant No.\,2000036.


\clearpage          

\thispagestyle{empty}
\setlength{\voffset}{25mm}
\begin{deluxetable}{lccccccccccr}
\tablecolumns{12}
\tabletypesize{\scriptsize}
\tablewidth{0truein}
\center
\tablecaption{Photometry of the High Latitude Sources
              \label{tab:targets}
              }
\tablehead{
\colhead{Star}&
\colhead{l}&
\colhead{b}&
\colhead{Spectral}&
\colhead{U}&
\colhead{B}&
\colhead{V}&
\colhead{J}&
\colhead{H}&
\colhead{K}

\\
\colhead{}&
\colhead{(deg)}&
\colhead{(deg)}&
\colhead{Type}&
\colhead{}&
\colhead{}&
\colhead{}&
\colhead{}&
\colhead{}&
\colhead{}
}
\startdata
HD\,19374&162.98&-34.21&B1.5V&5.20$\pm$0.03&5.99$\pm$0.02&6.11$\pm$0.02&6.341$\pm$0.019&6.472$\pm$0.029&6.470$\pm$0.018\\												
HD\,21483&158.87&-21.30&B3III&7.10$\pm$0.02&7.42$\pm$0.01&7.06$\pm$0.01&6.217$\pm$0.021&6.190$\pm$0.026&6.128$\pm$0.018\\												
HD\,24263&182.07&-34.87&B5V&5.30$\pm$0.02&5.73$\pm$0.01&5.67$\pm$0.01&5.421$\pm$0.029&5.436$\pm$0.024&5.446$\pm$0.024\\												
HD\,26571&172.42&-20.55&B8III&6.04$\pm$0.03&6.31$\pm$0.02&6.12$\pm$0.02&5.595$\pm$0.037&5.610$\pm$0.018&5.570$\pm$0.020\\												
HD\,28475&185.13&-25.09&B5V&6.47$\pm$0.02&6.84$\pm$0.02&6.77$\pm$0.02&6.465$\pm$0.024&6.514$\pm$0.033&6.466$\pm$0.017\\												
HD\,135485&347.31&35.46&B3V&7.59$\pm$0.03&8.13$\pm$0.02&8.20$\pm$0.02&8.258$\pm$0.017&8.322$\pm$0.033&8.371$\pm$0.021\\												
HD\,139094&343.03&23.20&B8IV&7.19$\pm$0.03&7.46$\pm$0.02&7.38$\pm$0.02&7.113$\pm$0.020&7.118$\pm$0.053&7.105$\pm$0.023\\												
HD\,140543&347.89&25.54&B0.5III&8.02$\pm$0.03&8.89$\pm$0.02&8.90$\pm$0.02&8.899$\pm$0.024&8.938$\pm$0.027&8.936$\pm$0.023\\												
HD\,141404&349.57&25.54&B9V&7.86$\pm$0.03&7.83$\pm$0.02&7.70$\pm$0.02&7.247$\pm$0.019&7.284$\pm$0.051&7.174$\pm$0.026\\												
HD\,141637&346.10&21.71&B2V&3.86$\pm$0.02&4.59$\pm$0.01&4.64$\pm$0.01&4.802$\pm$0.228&4.860$\pm$0.076&4.783$\pm$0.021\\												
HD\,142096&350.72&25.38&B3V&4.44$\pm$0.02&5.00$\pm$0.02&5.02$\pm$0.02&4.987$\pm$0.037&5.001$\pm$0.027&4.903$\pm$0.021\\												
HD\,142165&347.51&22.15&B5V&4.97$\pm$0.02&5.37$\pm$0.02&5.39$\pm$0.02&5.345$\pm$0.018&5.384$\pm$0.034&5.355$\pm$0.021\\												
HD\,142250&345.57&20.00&B7V&5.63$\pm$0.03&6.07$\pm$0.02&6.14$\pm$0.02&6.234$\pm$0.019&6.312$\pm$0.033&6.300$\pm$0.016\\												
HD\,142883&350.88&24.08&B3V&5.40$\pm$0.03&5.83$\pm$0.04&5.84$\pm$0.04&5.764$\pm$0.021&5.767$\pm$0.036&5.734$\pm$0.033\\												
HD\,143275&350.10&22.49&B0IV&1.29$\pm$0.02&2.20$\pm$0.01&2.32$\pm$0.01&2.410$\pm$0.308&2.432$\pm$0.214&2.427$\pm$0.278\\												
HD\,143567&350.87&22.68&B9V&7.18$\pm$0.03&7.27$\pm$0.02&7.19$\pm$0.02&6.928$\pm$0.021&6.946$\pm$0.029&6.903$\pm$0.020\\												
HD\,143600&350.37&22.13&B9V&7.38$\pm$0.03&7.43$\pm$0.02&7.33$\pm$0.02&7.037$\pm$0.018&7.038$\pm$0.018&6.970$\pm$0.029\\												
HD\,144470&352.75&22.77&B1V&3.11$\pm$0.02&3.92$\pm$0.01&3.96$\pm$0.01&4.160$\pm$0.282&4.188$\pm$0.228&4.009$\pm$0.036\\												
HD\,145502&354.61&22.70&B2V&3.41$\pm$0.02&4.05$\pm$0.01&4.01$\pm$0.01&3.970$\pm$0.236&3.795$\pm$0.226&3.878$\pm$0.288\\												
HD\,145554&354.57&22.56&B9V&7.71$\pm$0.03&7.79$\pm$0.02&7.65$\pm$0.02&7.174$\pm$0.024&7.171$\pm$0.057&7.071$\pm$0.029\\												
HD\,145631&354.70&22.54&B9V&7.70$\pm$0.03&7.73$\pm$0.02&7.58$\pm$0.02&7.051$\pm$0.018&7.050$\pm$0.049&6.941$\pm$0.024\\												
HD\,146029&352.78&20.23&B9V&7.40$\pm$0.03&7.45$\pm$0.02&7.38$\pm$0.02&7.073$\pm$0.023&7.080$\pm$0.034&7.027$\pm$0.026\\												
HD\,146416&353.98&20.60&B9V&6.48$\pm$0.03&6.63$\pm$0.02&6.61$\pm$0.02&6.514$\pm$0.024&6.515$\pm$0.038&6.485$\pm$0.027\\												
HD\,147009&355.00&20.89&B9V&8.59$\pm$0.03&8.35$\pm$0.02&8.06$\pm$0.02&7.255$\pm$0.020&7.138$\pm$0.034&7.087$\pm$0.024\\												
HD\,147010&355.49&20.88&B9II-III&7.31$\pm$0.03&7.56$\pm$0.02&7.40$\pm$0.02&6.754$\pm$0.021&6.750$\pm$0.036&6.677$\pm$0.018\\												
HD\,148184&357.93&20.68&B2V&3.95$\pm$0.05&4.70$\pm$0.02&4.42$\pm$0.02&3.404$\pm$0.288&3.149$\pm$0.194&2.885$\pm$0.280\\												
HD\,149757&6.28   &23.59&O9V&1.72$\pm$0.02&2.58$\pm$0.01&2.56$\pm$0.01&2.533$\pm$0.296&2.667$\pm$0.212&2.684$\pm$0.272\\												
HD\,154445&19.29 &22.93&B1.5V&5.16$\pm$0.02&5.80$\pm$0.02&5.64$\pm$0.02&5.311$\pm$0.007&5.351$\pm$0.057&5.290$\pm$0.023\\												
HD\,156247&22.73 &21.57&B5V&5.49$\pm$0.02&5.94$\pm$0.02&5.92$\pm$0.02&5.656$\pm$0.020&5.720$\pm$0.044&5.739$\pm$0.027\\												
HD\,203532&309.46&-31.74&B3IV-V&6.15$\pm$0.02&6.51$\pm$0.01&6.38$\pm$0.01&6.066$\pm$0.026&6.062$\pm$0.038&6.029$\pm$0.020\\												
HD\,210121&56.87 &-44.46&B3V&....&7.87$\pm$0.02&7.67$\pm$0.01&7.384$\pm$0.024&7.418$\pm$0.031&7.403$\pm$0.018\\												
HD\,211924&69.30 &-40.86&B5IV&4.88$\pm$0.01&5.34$\pm$0.01&5.37$\pm$0.01&5.444$\pm$0.037&5.496$\pm$0.040&5.452$\pm$0.016\\														
\enddata 
\\
\end{deluxetable}

\clearpage
\thispagestyle{empty}
\setlength{\voffset}{25mm}
\begin{deluxetable}{lcccccccccccccr}
\rotate
\tablecolumns{12}
\tabletypesize{\scriptsize}
\tablewidth{0truein}
\center
\tablecaption{Dereddening Parameters for the Comparison Stars
              \label{tab:deredd}
              }
\tablehead{
\colhead{Star}&
\colhead{l}&
\colhead{b}&
\colhead{Spectral}&
\colhead{$V$}&
\colhead{$E(B-V)$}&
\colhead{$A_V$}&
\colhead{$\cpone$}&
\colhead{$\cptwo$}&
\colhead{$\cpthree$}&
\colhead{$\cpfour$}&
\colhead{$\xo$}&
\colhead{$\gamma$}
\\
\colhead{}&
\colhead{(deg)}&
\colhead{(deg)}&
\colhead{Type}&
\colhead{(mag)}&
\colhead{(mag)}&
\colhead{(mag)}&
\colhead{}&
\colhead{}&
\colhead{}&
\colhead{}&
\colhead{($\mu$m$^{-1}$)}&
\colhead{($\mu$m$^{-1}$)}
}
\startdata
HD\,10205&132.92&-21.34&B8IV&4.94&0.02&0.062$\pm$0.01&0.84$\pm$0.31&0.26$\pm$0.06&1.07$\pm$0.20&0.15$\pm$0.03&4.62$\pm$0.21&0.97$\pm$0.12\\
HD\,25340&191.88&-37.81&B5V&5.28&0.01&0.031$\pm$0.01&0.85$\pm$0.33&0.27$\pm$0.05&1.10$\pm$0.21&0.14$\pm$0.02&4.61$\pm$0.22&0.99$\pm$0.12\\
HD\,17769&161.17&-38.71&B7V&5.49&0.04&0.124$\pm$0.02&0.79$\pm$0.23&0.27$\pm$0.07&1.12$\pm$0.24&0.14$\pm$0.03&4.60$\pm$0.19&0.99$\pm$0.11\\
HD\,25631&214.57&-45.74&B3V&6.45&0.04&0.124$\pm$0.01&0.74$\pm$0.24&0.28$\pm$0.07&1.06$\pm$0.19&0.13$\pm$0.02&4.61$\pm$0.23&0.96$\pm$0.10\\
HD\,31726&213.5&-31.51&B1V&6.15&0.05&0.155$\pm$0.01&0.80$\pm$0.28&0.27$\pm$0.06&1.03$\pm$0.22&0.15$\pm$0.03&4.60$\pm$0.22&0.95$\pm$0.10\\
HD\,36512&210.43&-20.98&B0V&4.62&0.04&0.124$\pm$0.01&0.90$\pm$0.36&0.25$\pm$0.06&1.12$\pm$0.21&0.17$\pm$0.04&4.61$\pm$0.25&1.00$\pm$0.11\\
HD\,38666&237.29&-27.1&O9V&5.17&0.03&0.093$\pm$0.01&0.81$\pm$0.29&0.27$\pm$0.06&1.02$\pm$0.22&0.14$\pm$0.02&4.60$\pm$0.21&0.95$\pm$0.13\\
HD\,149881&31.37&36.23&B0III&7.05&0.12&0.372$\pm$0.05&0.86$\pm$0.30&0.26$\pm$0.06&1.03$\pm$0.18&0.15$\pm$0.03&4.62$\pm$0.20&0.96$\pm$0.12\\
HD\,192273&325.78&-32.93&B2V&8.83&0.06&0.186$\pm$0.01&0.74$\pm$0.31&0.28$\pm$0.07&1.06$\pm$0.19&0.13$\pm$0.03&4.61$\pm$0.23&0.96$\pm$0.11\\
HD\,192907&110.38&22.5&B9III&4.39&0.03&0.093$\pm$0.01&0.84$\pm$0.27&0.26$\pm$0.06&1.14$\pm$0.25&0.14$\pm$0.03&4.60$\pm$0.20&1.00$\pm$0.14\\
HD\,212581&323.87&-45.87&B9V&4.48&0.04&0.124$\pm$0.01&0.80$\pm$0.18&0.27$\pm$0.05&1.09$\pm$0.17&0.15$\pm$0.05&4.60$\pm$0.15&0.98$\pm$0.11\\
HD\,220787&67.81&-64.4&B3III&8.31&0.03&0.093$\pm$0.02&0.84$\pm$0.31&0.26$\pm$0.05&1.06$\pm$0.20&0.15$\pm$0.03&4.62$\pm$0.19&0.97$\pm$0.12\\

\enddata 
\\
\end{deluxetable}

\thispagestyle{empty}
\setlength{\voffset}{15mm}
\begin{deluxetable}{lccccccccccccccr}
\rotate
\tablecolumns{15}
\tabletypesize{\tiny}
\tablewidth{0truein}
\center
\tablecaption{Extinction and FM90 Parameters
              for the HLC Sightlines
              \label{tab:extpara}
             }
\tablehead{
\colhead{Star}&
\colhead{l}&
\colhead{b}&
\colhead{Spectral}&
\colhead{Comparison}&
\colhead{$A_V$}&
\colhead{$E(B-V)$}&
\colhead{$R_V$}&
\colhead{$c_{1}^{\prime}$}&
\colhead{$c_{2}^{\prime}$}&
\colhead{$c_{3}^{\prime}$}&
\colhead{$c_{4}^{\prime}$}&
\colhead{$x_{0}$}&
\colhead{$\gamma$}
\\
\colhead{}&
\colhead{(deg)}&
\colhead{(deg)}&
\colhead{Type}&
\colhead{Star}&
\colhead{(mag)}&
\colhead{(mag)}&
\colhead{}&
\colhead{}&
\colhead{}&
\colhead{}&
\colhead{}&
\colhead{($\mu$m$^{-1}$)}&
\colhead{($\mu$m$^{-1}$)}
}
\startdata

HD 19374&162.98&-34.21&B1.5V&HD 31726&0.29$\pm$0.06&0.13$\pm$0.02&2.23$\pm$0.31&0.39$\pm$0.07&0.36$\pm$0.041&1.90$\pm$0.041&0.18$\pm$0.030&4.63$\pm$0.012&0.96$\pm$0.034\\
HD 21483&158.87&-21.30&B3III&HD 220787&1.60$\pm$0.08&0.56$\pm$0.02&2.85$\pm$0.10&0.86$\pm$0.11&0.29$\pm$0.025&1.16$\pm$0.025&0.22$\pm$0.028&4.60$\pm$0.006&1.16$\pm$0.051\\
HD 24263&182.07&-34.87&B5V&HD 25340   &0.70$\pm$0.08&0.22$\pm$0.02&3.20$\pm$0.19&1.86$\pm$0.22&0.03$\pm$0.002&1.28$\pm$0.002&0.20$\pm$0.021&4.57$\pm$0.004&0.95$\pm$0.033\\
HD 26571&172.42&-20.55&B8III&HD 10205  &0.91$\pm$0.10&0.29$\pm$0.03&3.15$\pm$0.12&0.43$\pm$0.09&0.36$\pm$0.052&1.02$\pm$0.052&0.26$\pm$0.040&4.60$\pm$0.015&0.99$\pm$0.021\\
HD 28475&185.13&-25.09&B5V&HD 25340   &0.81$\pm$0.08&0.23$\pm$0.02&3.51$\pm$0.16&1.27$\pm$0.19&0.13$\pm$0.010&1.27$\pm$0.010&0.06$\pm$0.010&4.59$\pm$0.014&0.89$\pm$0.040\\
HD 135485&347.31&35.46&B3V&HD 25631  &0.37$\pm$0.07&0.13$\pm$0.02&2.85$\pm$0.33&0.28$\pm$0.03&0.55$\pm$0.040&1.36$\pm$0.040&0.38$\pm$0.027&4.59$\pm$0.016&1.02$\pm$0.033\\
HD 139094&343.03&23.20&B8IV&HD 10205&0.62$\pm$0.12&0.19$\pm$0.03&3.27$\pm$0.36&0.60$\pm$0.13&0.38$\pm$0.050&0.91$\pm$0.050&0.27$\pm$0.040&4.59$\pm$0.013&0.97$\pm$0.029\\
HD 140543&347.89&25.54&B0.5III&HD 149881&0.69$\pm$0.08&0.27$\pm$0.02&2.55$\pm$0.23&0.78$\pm$0.18&0.26$\pm$0.041&1.30$\pm$0.041&0.14$\pm$0.030&4.59$\pm$0.012&0.92$\pm$0.031\\
HD 141404&349.57&25.54&B9V&HD 212581&0.79$\pm$0.11&0.20$\pm$0.03&3.94$\pm$0.21&0.94$\pm$0.19&0.23$\pm$0.029&0.95$\pm$0.029&0.26$\pm$0.032&4.64$\pm$0.015&0.94$\pm$0.027\\
HD 141637&346.10&21.71&B2V&HD 192273&0.52$\pm$0.08&0.19$\pm$0.02&2.75$\pm$0.28&0.58$\pm$0.10&0.32$\pm$0.041&1.35$\pm$0.040&0.07$\pm$0.020&4.58$\pm$0.017&0.98$\pm$0.039\\
HD 142096&350.72&25.38&B3V&HD 25631&0.62$\pm$0.08&0.18$\pm$0.02&3.42$\pm$0.23&1.26$\pm$0.16&0.16$\pm$0.012&1.26$\pm$0.012&0.07$\pm$0.010&4.59$\pm$0.013&1.04$\pm$0.032\\
HD 142165&347.51&22.15&B5V&HD 25340&0.48$\pm$0.07&0.18$\pm$0.02&2.65$\pm$0.20&1.52$\pm$0.28&0.10$\pm$0.010&1.19$\pm$0.010&0.11$\pm$0.010&4.61$\pm$0.015&1.03$\pm$0.030\\
HD 142250&345.57&20.00&B7V&HD 17769&0.14$\pm$0.06&0.06$\pm$0.02&2.28$\pm$0.67&0.08$\pm$0.01&0.55$\pm$0.040&1.83$\pm$0.041&0.14$\pm$0.021&4.61$\pm$0.018&0.88$\pm$0.026\\
HD 142883&350.88&24.08&B3V&HD 25631&0.69$\pm$0.10&0.18$\pm$0.02&3.81$\pm$0.27&0.98$\pm$0.23&0.26$\pm$0.030&1.26$\pm$0.030&0.15$\pm$0.020&4.58$\pm$0.013&0.93$\pm$0.025\\
HD 143275&350.10&22.49&B0IV&HD 149881&0.74$\pm$0.27&0.18$\pm$0.02&4.10$\pm$1.45&1.03$\pm$0.14&0.06$\pm$0.004&1.11$\pm$0.004&0.09$\pm$0.010&4.59$\pm$0.017&0.99$\pm$0.033\\
HD 143567&350.87&22.68&B9V&HD 212581&0.50$\pm$0.10&0.15$\pm$0.03&3.36$\pm$0.26&0.84$\pm$0.17&0.11$\pm$0.010&1.32$\pm$0.010&0.01$\pm$0.002&4.60$\pm$0.015&0.87$\pm$0.035\\
HD 143600&350.37&22.13&B9V&HD 212581&0.59$\pm$0.12&0.17$\pm$0.03&3.45$\pm$0.42&0.72$\pm$0.11&0.13$\pm$0.012&1.30$\pm$0.012&0.02$\pm$0.002&4.60$\pm$0.015&0.90$\pm$0.020\\
HD 144470&352.75&22.77&B1V&HD 31726&0.74$\pm$0.09&0.22$\pm$0.02&3.36$\pm$0.22&1.36$\pm$0.19&0.17$\pm$0.015&1.39$\pm$0.015&0.11$\pm$0.020&4.57$\pm$0.019&0.91$\pm$0.034\\
HD 145502&354.61&22.70&B2V&HD 192273&0.81$\pm$0.25&0.28$\pm$0.02&2.89$\pm$0.88&0.51$\pm$0.09&0.38$\pm$0.037&1.49$\pm$0.037&0.22$\pm$0.030&4.57$\pm$0.012&0.97$\pm$0.028\\
HD 145554&354.57&22.56&B9V&HD 212581&0.84$\pm$0.11&0.21$\pm$0.03&4.00$\pm$0.21&1.10$\pm$0.25&0.04$\pm$0.004&1.04$\pm$0.004&0.02$\pm$0.002&4.63$\pm$0.018&0.94$\pm$0.030\\
HD 145631&354.70&22.54&B9V&HD 212581&0.9$\pm$0.10&0.22$\pm$0.02&4.10$\pm$0.19&1.00$\pm$0.23&0.10$\pm$0.001&0.89$\pm$0.001&0.09$\pm$0.021&4.60$\pm$0.012&0.90$\pm$0.031\\
HD 146029&352.78&20.23&B9V&HD 212581&0.59$\pm$0.12&0.14$\pm$0.03&4.24$\pm$0.31&1.12$\pm$0.21&0.06$\pm$0.006&1.07$\pm$0.006&0.07$\pm$0.010&4.60$\pm$0.014&0.90$\pm$0.042\\
HD 146416&353.98&20.60&B9V&HD 212581&0.33$\pm$0.10&0.09$\pm$0.03&3.65$\pm$0.49&0.96$\pm$0.14&0.08$\pm$0.006&1.32$\pm$0.006&0.10$\pm$0.014&4.64$\pm$0.011&0.89$\pm$0.037\\
HD 147009&355.00&20.89&B9V&HD 212581&1.27$\pm$0.10&0.36$\pm$0.03&3.54$\pm$0.12&1.27$\pm$0.17&0.13$\pm$0.011&1.03$\pm$0.011&0.30$\pm$0.047&4.58$\pm$0.011&0.94$\pm$0.044\\
HD 147010&355.49&20.88&B9II-III&HD 192907&0.92$\pm$0.11&0.24$\pm$0.03&3.85$\pm$0.16&0.54$\pm$0.16&0.21$\pm$0.028&1.21$\pm$0.028&0.05$\pm$0.011&4.54$\pm$0.016&0.98$\pm$0.038\\
HD 148184&357.93&20.68&B2V&HD 192273&2.39$\pm$0.29&0.52$\pm$0.03&4.60$\pm$0.51&1.14$\pm$0.18&0.14$\pm$0.015&0.63$\pm$0.015&0.05$\pm$0.010&4.56$\pm$0.023&1.04$\pm$0.035\\
HD 149757&6.28&23.59&O9V&HD 38666&0.84$\pm$0.26&0.33$\pm$0.02&2.53$\pm$0.79&0.94$\pm$0.16&0.33$\pm$0.039&1.36$\pm$0.039&0.19$\pm$0.034&4.58$\pm$0.016&1.05$\pm$0.036\\
HD 154445&19.29&22.93&B1.5V&HD 31726&1.30$\pm$0.08&0.41$\pm$0.02&3.16$\pm$0.10&1.62$\pm$0.27&0.09$\pm$0.009&1.31$\pm$0.009&0.16$\pm$0.028&4.56$\pm$0.020&0.94$\pm$0.028\\
HD 156247&22.73&21.57&B5V&HD 25340&0.62$\pm$0.07&0.22$\pm$0.02&2.82$\pm$0.18&1.20$\pm$0.11&0.09$\pm$0.006&1.56$\pm$0.006&0.10$\pm$0.012&4.59$\pm$0.011&1.04$\pm$0.026\\
HD 203532&309.46&-31.74&B3IV-V&HD 25631&1.09$\pm$0.08&0.33$\pm$0.02&3.30$\pm$0.12&0.98$\pm$0.18&0.26$\pm$0.031&1.36$\pm$0.031&0.20$\pm$0.026&4.56$\pm$0.016&0.99$\pm$0.029\\
HD 210121&56.87&-44.46&B3V&HD 25631&0.84$\pm$0.07&0.40$\pm$0.03&2.10$\pm$0.10&-0.54$\pm$0.11&0.88$\pm$0.110&1.62$\pm$0.110&0.37$\pm$0.060&4.52$\pm$0.021&1.24$\pm$0.057\\
HD 211924&69.30&-40.86&B5IV&HD 25340&0.36$\pm$0.06&0.13$\pm$0.02&2.78$\pm$0.10&0.86$\pm$0.09&0.38$\pm$0.028&1.36$\pm$0.027&0.20$\pm$0.026&4.63$\pm$0.014&0.96$\pm$0.032\\

\enddata 
\\
\end{deluxetable}

\end{document}